\documentclass[opre,nonblindrev]{informs3} 

\DoubleSpacedXI 

\usepackage{endnotes}
\let\footnote=\endnote
\let\enotesize=\normalsize
\def\notesname{Endnotes}%
\def\makeenmark{$^{\theenmark}$}
\def\enoteformat{\rightskip0pt\leftskip0pt\parindent=1.75em
  \leavevmode\llap{\theenmark.\enskip}}

\usepackage{comment}
\usepackage{multirow}
\usepackage{makecell} 
\setcellgapes{5pt}
\usepackage{algorithm,algpseudocode}
\usepackage{graphicx}
\usepackage{enumitem}
\usepackage{booktabs}
\usepackage[caption=false]{subfig}
\usepackage{caption}
\graphicspath{{plots/}}
\usepackage[toc,page]{appendix}

\def\betavec{{\boldsymbol\beta}}
\def\gammavec{{\boldsymbol\gamma}}

\def\x{\mathbf{x}}
\def\z{\mathbf{z}}

\def\xt{\tilde{\x}}

\def\xUnlab1{{\xt^{1}}}

\def\w{\mathbf{w}}
\def\v{\mathbf{v}}

\newcommand{\Popu}{X}

\newcommand{\SetOutcomes}{Y} 
\newcommand{\Outcome}{y}
\newcommand{\predictor}{\mathcal{H}}
\newcommand{\protAttr}{A}
\newcommand{\unprotAttr}{Z}

\newcommand{\all}{\textnormal{ for all }}
\makeatletter
\newcommand*{\balancecolsandclearpage}{%
  \close@column@grid
  \clearpage
  \twocolumngrid
}

\makeatletter
\newcommand*{\rom}[1]{\expandafter\@slowromancap\romannumeral #1@}
\makeatother

\usepackage{amsfonts}
\usepackage{mathtools}

\def\x{\mathbf{x}}
\graphicspath{{plots/}}
\allowdisplaybreaks

\newtheorem{mydef}{Definition}

\usepackage{natbib}
 \bibpunct[, ]{(}{)}{,}{a}{}{,}%
 \def\bibfont{\small}%
\TheoremsNumberedThrough     
\ECRepeatTheorems

\EquationsNumberedThrough    

\begin{document}


\RUNAUTHOR{Bandi and Bertsimas}

\RUNTITLE{The Price of Diversity}

\TITLE{The Price of Diversity}

\ARTICLEAUTHORS{%
\AUTHOR{Hari Bandi}
\AFF{Operations Research Center, Massachusetts Institute of Technology, Cambridge, MA 02139, \EMAIL{hbandi@mit.edu}} 
\AUTHOR{Dimitris Bertsimas}
\AFF{Sloan School of Management and Operations Research Center, Massachusetts Institute of Technology, Cambridge, MA 02139, \EMAIL{dbertsim@mit.edu}}
} 

\ABSTRACT{%
Systemic bias with respect to gender, race and ethnicity, often unconscious, is prevalent in datasets involving choices among individuals. Consequently, society has found it challenging to alleviate bias and achieve diversity in a way that maintains meritocracy in such settings. We propose (a) a novel optimization approach based on optimally flipping outcome labels and training classification models simultaneously to discover changes to be made in the selection process so as to achieve diversity without significantly affecting meritocracy,  and (b) a novel implementation tool employing optimal classification trees to provide insights on which attributes of individuals lead to flipping of their labels, and to help make changes in the current selection processes in a manner understandable by human decision makers. We present case studies on three real-world datasets consisting of parole,  admissions to the bar and lending decisions,  and demonstrate that the price of diversity is low and sometimes negative, that is we can modify our selection processes in a way that enhances diversity without affecting meritocracy significantly, and sometimes improving it. 
}%



\maketitle

%


\section{Introduction}
In this paper, we consider bias with respect to gender, race and ethnicity, often unconscious, but prevalent in datasets involving choices made by people. Such examples include datasets consisting of choices made by human decision makers in college admissions, hiring, lending, or parole decisions that discriminate against people of certain color or gender (\cite{angwin2016machine}, \cite{wightman1998lsac}). Historical data aggregated in such settings maybe biased against certain demographic groups due to systemic bias, thereby leading to a skewed set of selected candidates. Our objective here is to alleviate bias and enhance \textsl{diversity} in such settings while maintaining \textsl{meritocracy} by placing constraints on some selected objective measures of \textsl{meritocracy}.

In the literature, \textsl{diversity} is defined as the practice of including  people from a range of different social and ethnic backgrounds, and of different genders and sexual orientations. In contrast, \textsl{meritocracy} is defined as the practice of selecting people,  based on achievement as opposed to wealth or membership in a special social class. Often in practice, these two goals are in conflict and we observe an inherent trade-off between them. Thus, we are in grave need of systematic approaches that alleviate systemic bias and enhance diversity without significantly affecting meritocracy.

In recent years due to the growing volume of applications in college admissions, hiring, lending among many others, machine learning models are being increasingly employed for such classification problems. Since the datasets used to train these models consist of historical data populated with choices made by people, systemic bias may be  concealed in them. Consequently, it has been shown that without appropriate intervention during training or evaluation, classification models trained on such datasets can be biased against certain groups of individuals (\cite{angwin2016machine}; \cite{hardt2016equality}). This is due to the fact that during the training process, bias present in the dataset becomes reinforced into the model (\cite{bolukbasi2016man}). Thus, bias alleviation in these machine learning models has become an increasingly important concern.

To address this problem, simple remedies, such as ignoring the protected attributes, e.g., gender, race, ethnicity, etc., are largely ineffective due to other factors being correlated with the protected attributes (see \cite{pedreshi2008discrimination}). Thus, the data can be inherently biased in possibly complex ways,  making it difficult to alleviate bias. Moreover, it is both unethical and illegal to design a system that makes decisions entirely  on the basis of protected demographic attributes. Consequently, classification models that are actively trained on such datasets consisting of both human-made and model-made choices can progressively become more biased over time through  feedback loops leading to  amplified bias against a certain subpopulation. Such feedback loops have been observed in predictive policing (\cite{lila2019identifying}), credit markets and other applications, and this problem of \textsl{disparity amplification} is a possibility in any deployed machine learning model that is trained on historical data. Therefore, it is critical for any machine learning model to actively identify and alleviate systemic biases, improving demographic diversity in predicted outcomes without significantly affecting meritocracy. 

To this end,  we propose a novel optimization model to optimally flip outcome labels,  and train logistic regression and soft-margin support vector machine models so as to enhance diversity without significantly affecting meritocracy and quantify the price of diversity in the selected cohort (refer to Figure \ref{fig:epsilon-delta-tradeoff-lawschool}).  We also propose a novel implementation tool using Optimal Classification Trees (OCTs)  \citep{bertsimas2017optimal, MLbook} to provide insights on attributes of individuals that lead to flipping of their labels and help in making changes to the current selection processes in a manner understandable by human decision makers.

\subsection{Related work}
The literature on fairness in classification and bias alleviation can be categorized into three main approaches: (a) pre-processing training data, (b) post-processing of a classifier's outputs (predictions),  and (c) a Lagrangian approach of transforming fairness constraints on the classifier to penalties and solving the resulting optimization problem.

\textbf{Pre-processing training data:} One approach to alleviate bias in a dataset involves pre-processing the training data to remove bias.  Some examples include \cite{calders2009building}; \cite{kamiran2012data}; \cite{vzliobaite2011handling}; \cite{feldman2015certifying}; \cite{beutel2019fairness}.  These approaches typically seek to transform the training dataset so that it can satisfy some specific notions of fairness when training classification models.  In practice,  this often leads to classifiers that still exhibit substantial bias and perform poorly compared to the state-of-art.

\textbf{Post-processing of classifier's outputs:} This approach of bias alleviation involves calibrating the output of a trained classifier. Some examples of previous work in this direction include \cite{doherty2012ends}; \cite{feldman2015computational}; \cite{hardt2016equality}.  \cite{pleiss2017fairness} showed that a deterministic solution is only compatible with a single fairness constraint and thus cannot be applied to a group of fairness constraints.  Furthermore,  calibration of the outputs after training a model can  lead to models with poor accuracy tradeoff.  In certain special cases, \cite{woodworth2017learning} showed that post-processing of a classifier's outputs can be provably suboptimal and in doing so, the resulting classifiers are incompatible with other notions of fairness (\cite{chouldechova2017fair}; \cite{kleinberg2016inherent}).

\textbf{Lagrangian approach:} There has been much recent work on enforcing fairness constraints on a classifier by transforming the resulting constrained optimization problem via the method of Lagrange multipliers.  \cite{zafar2017fairness} and \cite{goh2016satisfying} apply this approach to problems in convex setting, and (\cite{kearns2018preventing}; \cite{agarwal2018reductions}; \cite{cotter2019training}) apply this approach in the non-convex setting by framing the constrained optimization problem as a two-player game (also refer to \cite{edwards2015censoring}; \cite{corbett2017algorithmic}; \cite{narasimhan2018learning}).  Other works include adding penalties in the objective function to achieve various forms of fairness; (refer to \cite{donini2018empirical};\cite{komiyama2018nonconvex} work on linear models and kernel methods, respectively).  However, in most cases the fairness constraints are required to be relaxed in order to solve the corresponding optimization problem.  Also,  in a recent work \cite{cotter2019training},  it has been shown that training models can be difficult as the Lagrangian may not even have a solution to converge to.

Finally,  \cite{bertsimas2019robustclassification} propose a robust optimization based approach to train logistic regression and SVM models using noisy data both in the feature and label space.  They use an uncertainty set with a budget to quantify the noise in the features and allow flipping a subset of the labels (with a budget) to train a robust classification model.  They formulate a min-max formulation to achieve this.  We take inspiration from their approach to flip a subset of the labels so as to achieve diversity,  but in our approach we formulate a min-min optimization problem to flip a subset of the labels and simultaneously optimize for the parameters of the classification model to improve the overall fit.

\subsection{Contributions}
In this paper, we propose a novel optimization approach to train classification models on large datasets to alleviate bias and enhance diversity without significantly affecting meritocracy.  Specifically,  we develop algorithms for logistic regression and soft-margin Support Vector Machines (SVMs) models.  We also propose a novel implementation tool using OCTs to make changes to the current selection processes in a manner understandable by human decision makers.

We summarize our contributions in this paper below:
\begin{enumerate}
    \item  We propose a novel optimization approach based on simultaneously optimally flipping labels and  training classification models that alleviate systemic bias and achieve diversity without significantly affecting meritocracy,  thereby leading to classifiers that achieve demographic parity (similar rates of positive predictions across protected subgroups) without affecting meritocracy significantly.
    \item We present case studies on three real-world datasets involving parole,  admissions to the bar and lending decisions,  and demonstrate that the price of diversity is often low and sometimes negative, that is we can modify our selection processes in a way that enhances diversity without affecting meritocracy significantly, and sometimes improving it.
    \item  We propose a novel implementation tool employing OCTs to provide insights on which attributes of individuals lead to flipping of their labels,  and to help make changes in the current selection processes in a manner understandable by human decision makers.
\end{enumerate}

The rest of the paper is structured as follows. In Section \ref{sec:bias}, we propose an approach to both identify and quantify systemic biases in a dataset and present an objective to achieve diversity in the selection process. In Section \ref{sec:methodology}, we propose an optimization formulation to train classifiers with constraints on some selected objective measures of meritocracy that achieve an increase in diversity in its predictions.  We extend this formulation and develop specialized algorithms for logistic regression and soft-margin SVM models.  We present three case studies on real-world datasets involving parole decisions, admissions to the bar and lending in Section \ref{sec:results}.  In Section \ref{sec:implementation-tool}, we propose to use OCTs as an implementation tool to make changes in the current selection processes so as to enhance diversity,  and finally,  we summarize our key findings in Section \ref{sec:conclusions}.

\section{Quantifying systemic bias}\label{sec:bias}
In this section, we propose an approach to identify and quantify systemic bias present in a dataset against a certain subpopulation. We further review the notion of demographic parity and use it to achieve diversity in the predictions of a classifier.

For illustration purposes,  we present concepts used in the following section using a law students dataset (also used for a case study in Section \ref{sec:results}) that examines whether the bar exam taken by law students in the US prior to 2011 has been biased against   black students.  This dataset also contains anonymized historical information about the student who participated in this study including age,  LSAT scores,  law school GPA,  undergraduate GPA . 

Let $\Popu$, $\protAttr$ and $\unprotAttr$ represent the set of all students,  protected attributes (race) and remaining attributes (age,  undergraduate GPA, law school GPA, etc.),  respectively.  The students in the population are further divided into several groups depending on their race.  In this case,  we specifically focus on the set of White students ($\mathcal{W}$) and the set of Black students ($\mathcal{B}$). We  further  define a set $\mathcal{B}^{+}\subseteq \mathcal{B}$ consisting of all observations in $\mathcal{B}$ that were positively labeled and $\mathcal{B}^{-}\subseteq \mathcal{B}$   consisting of all observations in $\mathcal{B}$ that were negatively labeled.
Let $n_b = |\mathcal{B}|$,   $n_w = |\mathcal{W}|$, $p_b = |\mathcal{B}^+| $ and $  p_w = |\mathcal{W}^+|$. Based on the choices made by the bar council,  each of the students is assigned an outcome $\SetOutcomes = \{ -1,  +1\}$ representing either passing ($+1$) or failing ($-1$) the bar exam.  For any given student $\x_i \in \Popu$,  let $\Outcome_i$ be the true outcome (or label) to be predicted.  A (possibly randomized) predictor can be represented by a mapping $\predictor : \Popu \rightarrow \SetOutcomes$ from population $\Popu$ to the set of outcomes $\SetOutcomes$, such that $\predictor(\x_i)$ is the predicted outcome for individual $\x_i$.  We next  describe the notion of an $\alpha$-biased dataset and propose an approach to quantify bias present, if any, against a protected subpopulation. 

\subsection{$\alpha$-biased datasets}
Here,  we define a notion which we call \textsl{$\alpha$-biased dataset} to quantify disparity in outcomes between two subpopulations in a dataset.  We call a dataset $\alpha$-biased if the difference between the rates of positive observations among a pair of subgroups $\mathcal{W}$ and $\mathcal{B}$ based on a protected variable (in this case, race) is at least $\alpha$. We formalize the definition below.
\begin{mydef}[$\alpha$-biased dataset]
A dataset $\mathcal{X}=\lbrace (\x_i, y_i)|\ y_i\in\lbrace -1,1\rbrace\rbrace$ is said to be $\alpha$-biased with respect to a pair of subgroups $\mathcal{W}, \mathcal{B} \subseteq \mathcal{X}$ if
\begin{eqnarray}
\left\lvert \frac{\sum_{i\in \mathcal{W}}{\mathbb{I}(y_i=+1)}}{n_w} - \frac{\sum_{i\in \mathcal{B}}{\mathbb{I}(y_i=+1)}}{n_b}\right\rvert  \geq \alpha.
\nonumber
\end{eqnarray}
\end{mydef}
Note that this is a characteristic of the dataset and represents bias in terms of the rates of positive observations without accounting for discrepancies in the distributions of attributes $Z$, for example,  undergraduate GPA or law school GPA of the students belonging to the pair of  groups $\mathcal{W}$ and $\mathcal{B}$.  In other words,  the discrepancy might purely be due to discrepancy in merit among the two groups of students.  Therefore, although this definition points to a possibility of systemic bias against a certain subgroup of students, it does not provide any evidence of bias. 

\subsection{Demographic parity}
Here, we introduce  the notion of demographic parity for a classifier $\predictor$ and use it as an objective to alleviate bias and achieve diversity. Demographic parity 
 imposes the condition that a classifier $\predictor$ should predict a positive outcome for individuals across a given pair of subgroups with \textit{almost} equal frequency. A formal definition is given below.
\begin{mydef}(\textnormal{Demographic parity})\label{def:demographic-parity}
A classifier $\predictor : \Popu \rightarrow \lbrace -1, 1\rbrace$ achieves demographic parity with bias $\epsilon$ with respect to groups $\mathcal{W}, \mathcal{B} \subseteq \Popu$  if and only if 
\begin{equation}\label{eq:parity}
\left\lvert \frac{\sum_{i\in\mathcal{W}}{\mathbb{I}(\predictor(\x_i) = +1)}}{n_w} - \frac{\sum_{i\in\mathcal{B}}{\mathbb{I}(\predictor(\x_i) = +1)}}{n_b}\right\rvert  \leq \epsilon.\nonumber
\end{equation}
\end{mydef}

A drawback with imposing a demographic parity constraint is that the predictor is not stipulated to select the most meritorious students within a group as long as it maintains demographic parity. Therefore, in the next section we propose an optimization formulation to train classifiers that choose the most meritorious students within each subgroup while trying to achieve demographic parity, thereby alleviating bias and achieving diversity while simultaneously maintaining meritocracy.
 
\section{Bias Alleviation}\label{sec:methodology}
In this section, we propose a novel mixed integer optimization (MIO) formulation to train logistic regression and soft-margin SVMs given an $\alpha$-biased dataset that is biased against either of the subgroups $\mathcal{W}$ and $\mathcal{B}$.

\subsection{Model}
Given a labeled training dataset $\left\lbrace (\x_i, y_i),  ~y_i\in\{ -1,1\} ~\x_i \in \mathbb{R}^p \right\rbrace_{i\in [n]}$ there are various methods in the machine learning literature to build classification models. However,  most of these methods are not designed to protect against systemic bias present in a dataset.

Here, we consider binary classification methods including logistic regression and SVMs,  and propose algorithms to train these models on an $\alpha$-biased dataset. We assume without loss of generality that the population can be grouped into two subgroups $\mathcal{W}$ and $\mathcal{B}$ based on some protected attribute  (in this case race) with one of the groups being discriminated against. 

A general classification algorithm proposes to learn parameters $\theta \in \Theta$ by minimizing a loss function $\ell(y,\x,\theta)$ through empirical risk minimization as follows:
\begin{eqnarray}\label{general-classification-problem}
\hat{\theta} = &&\argmin_{\theta\in\Theta}{\frac{1}{n}\sum_{i=1}^n{\ell(y_i,\x_i, \theta)}}.
\end{eqnarray}

The primary goal of our approach is to ensure demographic parity by allowing the model to flip some of the outcome labels while simultaneously optimizing the parameters of the classification model in order to improve the overall fit.  We propose a MIO problem that achieves this by introducing a binary variable $z_i \in \lbrace 0,1\rbrace, ~i\in [n]$ to decide whether the label for the $i^{th}$ observation is to be flipped (motivated by \cite{bertsimas2019robustclassification}).  If the original label is $y_i\in \lbrace -1,1\rbrace$, then the corresponding modified outcome label would be  $\tilde{y}_i = y_i(1-2z_i)$. Consequently, we define a set of $n$ binary variables that flip  a  proportion  $\tau_w$ of labels in $\mathcal{W}$ and a proportion $\tau_b$ of labels in $\mathcal{B}$ given by:
\begin{eqnarray}
\mathcal{Z}_{\tau_w,\tau_b} = \left\lbrace \z \in \lbrace 0,1\rbrace^n: \sum_{i\in \mathcal{W}}{z_i} = \lceil  \tau_w\cdot n_w \rceil , ~ \sum_{i\in \mathcal{B}} {z_i} = \lceil \tau_b\cdot n_b \rceil \right\rbrace.
\end{eqnarray}

Therefore, we would like to train a classification model on a modified dataset (through flipping labels) that helps the model to achieve lower loss. We can do both of these steps simultaneously by solving a ``min-min'' optimization problem that decides which labels to flip while learning a classifier. We propose to solve the following optimization problem:
\begin{eqnarray}\label{general-problem}
\hat{\theta} = &&\argmin_{\theta\in\Theta}{\min_{z\in \mathcal{Z}_{\tau_w,\tau_b}}{\frac{1}{n}\sum_{i=1}^n{\ell(y_i(1-2z_i),\x_i, \theta)}}}.
\end{eqnarray}
Next, we present special cases of Problem (\ref{general-problem}) to train logistic regression AND soft-margin SVMs with constraints on selected objective measures of meritocracy in the following sections.

\subsection{Logistic Regression} 
In this section,  we develop an optimization formulation for logistic regression model using Problem (\ref{general-problem}).  A logistic regression model assumes that the target variable $Y$ follows a Bernoulli distribution with probability given by input $\x \in \mathbb{R}^p$ and parameters $(\beta_0, \betavec) \in \mathbb{R}^{p+1}$. The optimization problem to minimize the loss function for a logistic regression model is given as follows:
\begin{eqnarray}
\min_{\beta_0,\betavec}~~~{\sum_{i=1}^n{\log\left(1+e^{-y_i(\betavec^\top \x_i + \beta_0)}\right)}}\nonumber.
\end{eqnarray}

Motivated by Problem (\ref{general-problem}),  we formulate a ``min-min'' optimization problem with binary variables $\z \in \{0, 1 \}^n$ to flip outcomes (labels) while placing constraints on some selected objective measures of meritocracy $\mathcal{M}$,  for example law school GPA,  LSAT scores, etc.  For each covariate in $\mathcal{M}$, the meritocracy constraints restrict the mean z-score of positively labeled observations to change at most by $\delta$ (meritocracy tolerance) in the modified dataset.  

Without loss of generality,  assume that all covariates are standardized with a mean of zero and unit standard deviation.  Let $\bar{x}_j$ denote the mean z-score of attribute $j\in\mathcal{M}$ among the postively labeled observations in the original dataset, i.e.,  $\displaystyle \bar{x}_j =  \frac{\sum_{i=1}^n{x_{ij} \left(y_i+1 \right)}}{2 (p_b + p_w) } $, note that since $y_i \in \{-1, +1\}$,  $(y_i + 1)/2$ is 1 if $y_i = +1$ and 0 if $y_i = -1$.  Similarly, the corresponding mean z-score of positively labeled observations in the modified dataset is given by $\displaystyle  \frac{\sum_{i=1}^n{x_{ij} \left(y_i (1-2 z_i)+1 \right)}}{\sum_{i=1}^n{\left(y_i (1-2 z_i)+1 \right) }}  $.  

Putting all of this together,  we propose to solve the following optimization problem:
\begin{align}
\min_{\beta_0,\betavec, \z\in\mathcal{Z}_{\tau_w,\tau_b}}~~~ &{ \sum_{i=1}^n{\log\left(1+e^{-y_i(1-2z_i)(\betavec^\top \x_i + \beta_0)}\right)}}& \label{logistic-main} \\
\text{s.t}~~~~& \left\lvert \frac{\sum_{i=1}^n{x_{ij} \left(y_i (1-2 z_i)+1 \right)}}{\sum_{i=1}^n{\left(y_i (1-2 z_i)+1 \right) }} - \bar{x}_{j} \right\rvert  \leq \delta, ~ j \in \mathcal{M}.  & \nonumber
\end{align}

Note that the above optimization problem can be reformulated by linearizing the product terms $\gamma_{i,j} = z_i \cdot  \beta_j, ~i\in [n], ~j\in [p]$ in the objective function with big-$M$ constraints as follows:
\begin{align}
\min_{\beta_0, \betavec, \gammavec, \z}~~~~& {f(\betavec, \gammavec) := \sum_{i=1}^n{\log\left(1+e^{-y_i(\betavec^\top \x_i + \beta_0)+2y_i(\gammavec_i^\top \x_i + \gamma_{i,0})}\right)}}& \label{logistic-bigM} \\
\text{s.t}~~~~& \sum_{i\in\mathcal{W}}{z_i}= \lceil \tau_w\cdot n_w\rceil ,& \nonumber\\
& \sum_{i\in\mathcal{B}}{z_i} = \lceil \tau_b\cdot n_b\rceil ,& \nonumber\\
&-z_i M_{j} \leq \gamma_{i,j}  \leq z_i M_{j},\ i\in [n], j\in [p],& \nonumber \\
&-(1-z_i)M_{j}\leq \gamma_{i,j} -\beta_j  \leq  (1-z_i)M_{j},\ i\in [n], j\in [p],& \nonumber \\
& \bar{x}_{j} - \delta \leq \frac{\sum_{i=1}^n{x_{ij} \left(y_i (1-2 z_i)+1 \right)}}{\sum_{i=1}^n{\left(y_i (1-2 z_i)+1 \right) }}  \leq \bar{x}_{j} + \delta, ~ j \in \mathcal{M},  & \nonumber \\
&\sum_{i\in \mathcal{W}} \gamma_{i,j}  = \lceil  \tau_w \cdot  n_w\rceil  \cdot \beta_j,\ j\in [p],& \nonumber \\
&\sum_{i\in \mathcal{B}} \gamma_{i,j}   = \lceil  \tau_b  \cdot  n_b\rceil  \cdot \beta_j,\ j\in [p],&\nonumber \\
&z_i \in \lbrace 0,1\rbrace, ~i\in [n]. & \nonumber
\end{align}
The meritocracy constraints can be linearized as follows:
$$
(\bar{x}_{j} - \delta) \sum_{i=1}^n{\left(y_i (1-2 z_i)+1 \right) } \leq \sum_{i=1}^n{x_{ij} \left(y_i (1-2 z_i)+1 \right)}  \leq (\bar{x}_{j} + \delta) \sum_{i=1}^n{\left(y_i (1-2 z_i)+1 \right) }, ~ j \in \mathcal{M}.
$$
We note that Problem (\ref{logistic-bigM}) is a mixed integer nonlinear optimization (MINLO) problem and that the objective function is convex in $(\betavec, \gammavec)$. We propose to solve Problem (\ref{logistic-bigM})  using an outer approximation (OA) method which iteratively refines and optimizes a piecewise linear approximation of $f(\betavec, \gammavec)$ (\cite{duran1986outer}).  The OA approach alternates between solving a mixed integer linear optimization problem (approximation to Problem (\ref{logistic-bigM})) and a nonlinear optimization problem with binary variables ($\z$) fixed to the solution from the previous iteration.  

We start by formulating a mixed integer linear optimization problem as an approximation to problem (\ref{logistic-bigM}) and keep adding linear estimators to $f(\betavec, \gammavec)$ dynamically.   At iteration t,  let $\left(\betavec^t = \betavec^\star(\z^t), \gammavec^t = \gammavec^\star(\z^t)\right)$ be the solution of the nonlinear optimization problem with binary variables fixed at $\z = \z^t$ ($\z^t$ is a solution of the mixed integer linear approximation to Problem (\ref{logistic-bigM}) at iteration $t$).  At iteration $t+1$,  a linear approximation to the objective function $f(\betavec, \gammavec)$ at $\left(\betavec^t, \gammavec^t\right)$ is added to the mixed integer linear optimization problem as a cut using the function value $f\left(\betavec^t, \gammavec^t\right)$ and gradient $\nabla f\left(\betavec^t, \gammavec^t\right)$ as given below,
$$f\left(\betavec, \gammavec\right) \geq f\left(\betavec^t, \gammavec^t\right) + \nabla f\left(\betavec^t, \gammavec^t\right)^\top \left(\betavec - \betavec^t, \gammavec - \gammavec^t\right).$$

Let $\displaystyle q_i = \left(1+e^{-y_i(\betavec^\top \x_i + \beta_0)+2y_i(\gammavec_i^\top \x_i + \gamma_{i,0})}\right)^{-1},~i\in [n]$,  the gradient $\nabla f(\betavec, \gammavec)$ is given as follows,
\begin{eqnarray}
&\displaystyle \frac{\partial f}{\partial \betavec} ~=~ - \sum_{i=1}^n{y_i\cdot q_i\cdot \x_i }, ~~~~~ \frac{\partial f}{\partial \gammavec_i} ~=~ 2y_i \cdot q_i\cdot \x_i , \ i\in [n], & \nonumber\\
&\displaystyle \frac{\partial f}{\partial \beta_0} ~=~ -\sum_{i=1}^n{y_i\cdot q_i } , ~~~~\frac{\partial f}{\partial \gamma_{i,0}} ~=~ 2y_i\cdot q_i,\ i\in [n]. &\nonumber
\end{eqnarray}

The solution approach proceeds by formulating a reduced master problem (RMP) at iteration $t+1$ as follows:
\begin{align}
\min_{\eta, \beta_0, \betavec, \gammavec, \z}~~~~& {\eta}& \label{RMP} \\
\text{s.t}~~~~& \eta \geq  f\left(\betavec^k, \gammavec^k\right) + \nabla f\left(\betavec^k, \gammavec^k\right)^\top \left(\betavec - \betavec^k, \gammavec - \gammavec^k\right), k \in [t] ,& \nonumber\\
& \sum_{i\in\mathcal{W}}{z_i}= \lceil \tau_w\cdot n_w\rceil ,& \nonumber\\
& \sum_{i\in\mathcal{B}}{z_i} = \lceil \tau_b\cdot n_b\rceil ,& \nonumber\\
&-z_i M_{j} \leq \gamma_{i,j}  \leq z_i M_{j},\ i\in [n], ~j\in [p],& \nonumber \\
&-(1-z_i)M_{j}\leq \gamma_{i,j} -\beta_j  \leq  (1-z_i)M_{j},\ i\in [n], ~j\in [p],& \nonumber \\
& \bar{x}_{j} - \delta \leq \frac{\sum_{i=1}^n{x_{ij} \left(y_i (1-2 z_i)+1 \right)}}{\sum_{i=1}^n{\left(y_i (1-2 z_i)+1 \right) }}  \leq \bar{x}_{j} + \delta, ~ j \in \mathcal{M},  & \nonumber \\
&\sum_{i\in \mathcal{W}} \gamma_{i,j}  = \lceil  \tau_w \cdot  n_w\rceil  \cdot \beta_j,\ j\in [p],& \nonumber \\
&\sum_{i\in \mathcal{B}} \gamma_{i,j}   = \lceil  \tau_b  \cdot  n_b\rceil  \cdot \beta_j,\ j\in [p],&\nonumber \\
&z_i \in \lbrace 0,1\rbrace, ~i\in [n]. & \nonumber
\end{align}

Note that the RMP is an approximation to Problem (\ref{logistic-bigM}) and we continuously refine it by adding linear approximations to the objective function in Problem (\ref{logistic-bigM}).  Solving the RMP at iteration $t+1$,  we obtain a solution $\z^{t+1}$ and use this solution to formulate a nonlinear optimization problem by fixing the binary variables $\z = \z^{t+1}$ in Problem (\ref{logistic-bigM}).  Note that by fixing binary variables $\z=\z^{t+1}$, we can update the outcome labels in Problem (\ref{logistic-main}) by setting $\tilde{y}_i = y_i\cdot (1-2\cdot z_i^{t+1}), ~ i\in [n]$,  and the meritocracy constraints are already satisfied as $\z^{t+1}$ is a solution of the RMP.  Therefore,  the nonlinear optimization problem (\ref{logistic-main}) at iteration $t+1$ reduces to a standard logistic regression problem with updated outcome labels $\tilde{y}$  as given below:
\begin{eqnarray}
\beta_0^{t+1},~ \betavec^{t+1} ~=~ \argmin_{\beta_0,\betavec}{ \sum_{i=1}^n{\log\left(1+e^{-\tilde{y}_i(\betavec^\top \x_i + \beta_0)}\right)}} .\nonumber
\end{eqnarray}

This problem can be solved efficiently using an off-the-shelf package (in our computational results we use the \texttt{glmnet} package for Julia v1.5).  The solutions $\left(\gamma_0^{t+1}, ~\gammavec^{t+1}\right)$ can be computed as follows: $\gamma_{0,i}^{t+1} = z_i^{t+1}\cdot \beta_0^{t+1} , ~ \gammavec_i^{t+1} = z_i^{t+1} \cdot \betavec^{t+1},~ i \in [n]$.  Using $\left(\beta_0^{t+1}, ~ \gamma_0^{t+1}, ~\betavec^{t+1}, ~\gammavec^{t+1}\right)$,  we can obtain $\z^{t+2}$ by solving an updated RMP with a new cut,  and we can compute $f(\betavec^{t+1}, \gammavec^{t+1})$ as a non-decreasing sequence of estimators that converges to the optimal value $f(\betavec^\star, \gammavec^\star)$ in finite number of iterations (as the outer-approximation approach never visits a point twice) (for a proof of convergence in finite number of iterations,  yet exponential in the worst case,  see \cite{fletcher1994solving}).  Moreover,  by dynamically generating contraints (using \texttt{lazy constraint callbacks}) we can avoid solving a different mixed integer optimization problem at each iteration.  The global minimum of Problem (\ref{logistic-bigM}) is reached when the objective function of the RMP is within some pre-specified tolerance of the objective function of the logistic regression problem with fixed outcome labels.

\subsection{Soft-Margin Support Vector Machines}
In this section,  we develop an optimization formulation for a soft-margin SVM model based on Problem (\ref{general-problem}). A soft-margin SVM is a variation of SVMs that allows observations to be incorrectly classified and penalizes them using a parameter C while maximizing the margin of the classifier. The corresponding optimization problem solves for parameters $\left(b, \w\right)$ that characterize the separating hyperplane of the SVM model. The resulting optimization problem is given as follows:
\begin{align}
\min_{b, \w}~~~~&\displaystyle {\frac{1}{2} \lVert \w\rVert_2^2 + C\sum_{i=1}^n{\xi_i}}&\nonumber \\
\text{s.t}~~~~&y_i(\w^\top \x_i - b) \geq 1 - \xi_i, ~ i \in [n], & \nonumber\\
& \xi_i \geq 0, ~ i\in [n]. \nonumber
\end{align}

Similar to logistic regression model in Problem (\ref{logistic-main}),  we formulate a ``min-min''  optimization problem for a soft-margin SVM with meritocracy constraints as given below,
\begin{align}
\min_{b, \w, ~\z\in \mathcal{Z}_{\tau_w,\tau_b}}~~~~&  {\frac{1}{2} \lVert \w\rVert_2^2 + C\sum_{i=1}^n{\xi_i}}& \label{SVM-main} \\
\text{s.t}~~~~&y_i(1-2z_i)(\w^\top \x_i - b) \geq 1- \xi_i,& \nonumber\\
& \left\lvert \frac{\sum_{i=1}^n{x_{ij} \left(y_i (1-2 z_i)+1 \right)}}{\sum_{i=1}^n{\left(y_i (1-2 z_i)+1 \right) }} - \bar{x}_{j} \right\rvert  \leq \delta, ~ j \in \mathcal{M}, & \nonumber \\
& \xi_i \geq 0, ~ i\in [n]. \nonumber
\end{align}
The above optimization problem can be reformulated by linearizing the product terms $v_{i,j} = z_i\cdot w_j, ~i\in [n], j\in [p]$ and $d_i = z_i\cdot b, ~i\in [n]$  using big-$M$ constraints as follows:
\begin{align}
\min_{b,d, \z,\w, \v}~~~~&\displaystyle  { \frac{1}{2} \lVert \w\rVert_2^2 + C\sum_{i=1}^n{\xi_i}}& \label{SVM-bigM} \\
\text{s.t}~~~~& \sum_{i\in\mathcal{W}}{z_i} =\lceil  \tau_w\cdot n_w\rceil ,& \nonumber\\
& \sum_{i\in\mathcal{B}}{z_i} = \lceil \tau_b\cdot n_b\rceil ,& \nonumber\\
& y_i(\w^\top \x_i - b) - 2y_i(\v_i^\top \x_i - d_i) \geq 1- \xi_i, ~i\in  [n], &\nonumber \\
&-z_i M_{j} \leq v_{i,j}  \leq z_i M_{j},~ i\in [n], ~j\in [p],& \nonumber \\
&-(1-z_i)M_{j} \leq v_{i,j} - w_j \leq  (1-z_i)M_{j},~ i\in [n], ~j\in [p],& \nonumber \\
& \bar{x}_{j} - \delta \leq \frac{\sum_{i=1}^n{x_{ij} \left(y_i (1-2 z_i)+1 \right)}}{\sum_{i=1}^n{\left(y_i (1-2 z_i)+1 \right) }}  \leq \bar{x}_{j} + \delta, ~ j \in \mathcal{M},  & \nonumber \\
&\sum_{i\in\mathcal{W}}{v_{i,j}}  =\lceil  \tau_w\cdot n_w \rceil  \cdot w_j,~ j\in [p],& \nonumber\\
&\sum_{i\in\mathcal{B}}{v_{i,j}}  = \lceil \tau_b\cdot n_b\rceil  \cdot w_j,~ j\in [p],&\nonumber \\
&z_i \in \lbrace 0,1\rbrace, ~i\in [n]. & \nonumber
\end{align}

Note that similar to logistic regression problem, the meritocracy constraints can be linearized.  Also, note that Problem (\ref{SVM-bigM}) is a Mixed-integer Quadratic Optimization (MIQO) problem and can be solved using an off-the-shelf MIQO solver.  In the subsequent section, we present an algorithm that trains logistic regression and SVM models that ensure $\epsilon$-demographic parity ($\epsilon \geq 0$) on a given $\alpha$-biased dataset.

\subsection{$\epsilon$-demographic parity classifier}
Here, we present an algorithm to train classifiers that achieves $\epsilon$-demographic parity among a given pair of protected groups $\mathcal{W}$ and $\mathcal{B}$ while selecting the most meritorious individuals in each of the subgroups on a $\alpha$-biased dataset. 

For a classifier to achieve $\epsilon$-demographic parity,  the parameters $\tau_w$ and $\tau_b$ need to be chosen in a way so as to ensure that the resulting classifier satisifies Definition \ref{def:demographic-parity}.  Observe that the number of labels allowed to be flipped for the pair of subgroups $\mathcal{W}$ and $\mathcal{B}$ are restricted by the parameters $\tau_w$ and $\tau_b$.  In order to achieve $\epsilon$-demographic parity,  our objective is twofold,  (1) match the rates of positive observations in both subgroups in the modified dataset, and (2) make sure that the total number of selected candidates is constant, i.e.,  the total number of positively labeled observations is constant after flipping outcome labels.  Without loss of generality,  assume that the rate of positively labeled observations is greater in $\mathcal{W}$ among the two subgroups, i.e., $ \displaystyle \frac{p_w}{n_w} > \frac{p_b}{n_b}$.  Since our objective is to decrease the difference between the rates of positive observations from $\alpha$ to $\epsilon$,  we decrease the rates of positive observations in $\mathcal{W}$ by $\tau_w$ and increase in $\mathcal{B}$ by $\tau_b$, and also make sure that the number of labels flipped among the two subgroups is the same. Therefore,  we can derive the following two equations to compute $\tau_w$ and $\tau_b$:
\begin{eqnarray}\label{tau-eqn}
 \displaystyle\left(\frac{p_w}{n_w} - \tau_w\right) - \left(\frac{p_b}{n_b} + \tau_b\right) =  \epsilon,  ~~
\tau_b\cdot n_b=\tau_w\cdot n_w.  
\end{eqnarray}

Solving the two equations above,  the values of $\tau_w$ and $\tau_b$ are given below:
\begin{align}
&\tau_w =  \frac{n_b \cdot p_w - p_b \cdot n_w - n_w \cdot n_b \cdot \epsilon}{n_w \cdot (n_w + n_b)}, &\nonumber \\
&\tau_b =  \frac{n_b \cdot p_w - p_b \cdot n_w - n_w \cdot n_b \cdot \epsilon}{n_b \cdot (n_w + n_b)}. &\nonumber
\end{align}

In order to train classifiers that achieve $\epsilon$-demographic parity,  we first standardize all unprotected covariates by subgroup,  i.e., we standardize covariates for the observations in the subgroups $\mathcal{W}$ and $\mathcal{B}$ separately.  Using the parameters $(\tau_w$,  $\tau_b)$,  we solve optimization Problem (\ref{logistic-bigM}) for logistic regression and Problem (\ref{SVM-bigM}) for soft-margin SVM to train classifiers that achieve $\epsilon$-demographic parity.  We present an algorithm to debias a given dataset and train classifiers that achieve $\epsilon$-demographic parity with respect to a pair of protected groups $\mathcal{W}$ and $\mathcal{B}$ in Algorithm \ref{alg:debiasing}.  
\begin{algorithm}[!h]
\caption{Algorithm for debiasing a dataset and training a classifier that achieves $\epsilon$-demographic parity.}\label{alg:debiasing}
{\bf Input:} Dataset $\mathcal{X}$,  demographic parity tolerance $\epsilon$ and meritocracy tolerance $\delta$.

{\bf Output:} Debiased classifier $\predictor$ with $\epsilon$-demographic parity.
\begin{algorithmic}[1]
\Procedure{DebiasDataset}{$\mathcal{X}$, $\epsilon$,  $\delta$}
\State Standardize unprotected covariates $\unprotAttr$ separately for each of the  subgroups $\mathcal{W}$ and $\mathcal{B}$, i.e.,  set $\displaystyle\tilde{x}_z = \frac{x_z - \mu_z^w}{\sigma_z^w} \all z \in \unprotAttr,  x \in \mathcal{W}$ and $\displaystyle\tilde{x}_z = \frac{x_z - \mu_z^b}{\sigma_z^b} \all z \in \unprotAttr,  x \in \mathcal{B}$ .   
\State Set parameters $\tau_w,~\tau_b $ to ensure $\epsilon$-demographic parity as follows:
$$\tau_w = \frac{n_b \cdot p_w - p_b \cdot n_w - n_w \cdot n_b \cdot \epsilon}{n_w \cdot (n_w + n_b)}, ~ \tau_b =  \frac{n_b \cdot p_w - p_b \cdot n_w - n_w \cdot n_b \cdot \epsilon}{n_b \cdot (n_w + n_b)}.$$
\State Train classifier $\predictor$ by solving optimization Problem (\ref{logistic-bigM}) for logistic regression and Problem (\ref{SVM-bigM}) for soft-margin SVM with demographic parity tolerance $\epsilon$ and meritocracy tolerance $\delta$.  
\EndProcedure
\end{algorithmic}
\end{algorithm}  

In the rest of the paper, we will refer to a logistic regression model trained using Algorithm \ref{alg:debiasing} as DP-LR (logistic regression model that achieves $\epsilon$-demographic parity), and similarly, an SVM model trained using Algorithm \ref{alg:debiasing} as DP-SVM.

\section{Computational Results}\label{sec:results}
In this section, we present computational results from case studies on three real-world datasets: (1) Bar admissions (law students dataset), (2) Recidivism (COMPAS dataset), and (3) Lending (credit default dataset) using Algorithm \ref{alg:debiasing} for both logistic regression and soft-margin SVM models.  We quantify the price of diversity enabled by the classification models trained using our approach in terms of the change in the mean of some selected objective measures of meritocracy among the positively predicted observations from the mean of the same attributes among the positively labeled observations in the original dataset. 

\subsection{Datasets}

We present case studies on three real-world datasets pertaining to admissions, recidivism and lending.  We briefly describe the datasets below:
\begin{itemize}
    \item \textbf{Law Students dataset (\cite{wightman1998lsac}):} This dataset originates from a study by \cite{wightman1998lsac} that examined whether the bar exam (taken by law students in the US) is biased against ethnic minorities.  The dataset contains anonymized historical information about the law students who participated in this study including age,  LSAT scores,  first year law school GPA (z-scores),  cumulative law school GPA (z-scores) and undergraduate GPA. 
    
    Note that the bar exam is administered by the individual states except two states in the US and the exam has been standardized since 2011.  Prior to 2011, the bar passage decision was based on subjective measures that required: (1) a multi-state bar examination (which was not a standardized exam in 1991),  (2) a multi-state essay examination,  (3) a multi-state performance test (a lawyering task, such as a memo or a brief),  and (4) undergoing a character and fitness examination.  This dataset originated from a longitudinal bar passage study from 1991-1997 and we use this dataset as an example to illustrate our methodology and its benefits had this methodology been available at the time.

    \item \textbf{ProPublica's COMPAS dataset (\cite{angwin2016machine}):} In this dataset, the task is to predict a risk score (high/low) for recidivism based on criminal history,  prison time and demographics. The historical dataset contains information about the defendants' age,  race,  sex,  number of prior convictions (prior counts) and COMPAS assigned risk scores (decile scores).  Here,  the protected attribute is race.  
    
    \item \textbf{Credit default dataset (\cite{yeh2009comparisons}):}  This dataset contains data from Taiwanese credit card users,  and the goal is to predict whether a given individual will default on payments for the subsequent month.  The dataset also contains attributes of individuals such as their approved credit limit (New Taiwan dollar),  gender,  education,  marital status,  history of payment,  bill and payment amounts (NT dollar).  The protected attribute in this case study is gender.
\end{itemize}

\subsection{Law school students dataset}
In this case study, we consider outcomes of bar exam in the US from a study conducted by Law School Admission Council (LSAC) (\cite{wightman1998lsac}). It contains information collected over a period of six years from 1991-1997 about law students taking the bar exam across various law schools in the US.  The information includes age, gender, race, ethnicity, grades and part time/full time status.  The goal is to determine whether or not an individual passed the bar exam.  We investigate bias with respect to both gender and race in this example. 

After preprocessing the data by removing instances that had missing values and those belonging to other ethnicity groups (neither black nor white) we were left with $20,738$ observations each with 11 features.  The resulting dataset is $\alpha$-biased with respect to race with  $\alpha = 0.3034$ and with respect to gender with $\alpha = 0.0211$.

\begin{figure}[!htbp]
\centering
\caption{Distributions of LSAT scores,  law school and undergraduate GPA candidates who passed the bar exam grouped by race and gender in the LSAC dataset which is $\alpha$-biased with respect to race with  $\alpha = 0.3034$ and with respect to gender with $\alpha = 0.0211$.}
\label{fig:LSAC-discrepancy}
\subfloat[Disparity across race labels.]{
\centering
\includegraphics[width=0.48\linewidth]{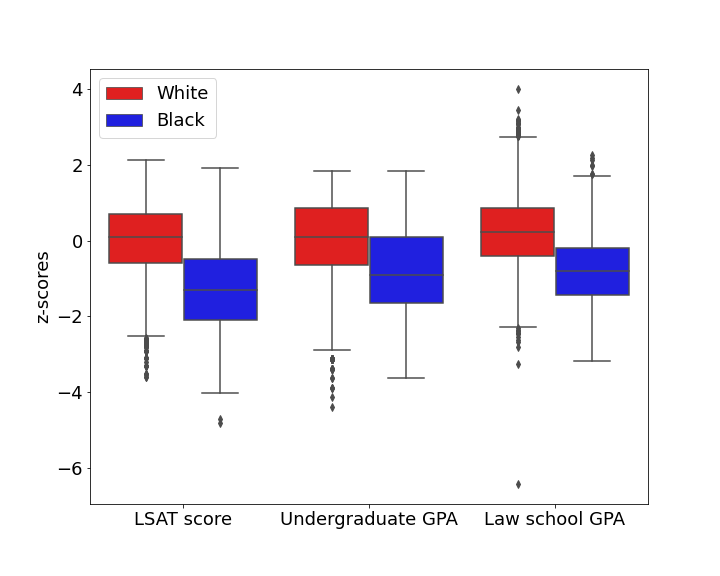}
}
\subfloat[Parity across gender labels.]{
\centering
\includegraphics[width=0.48\linewidth]{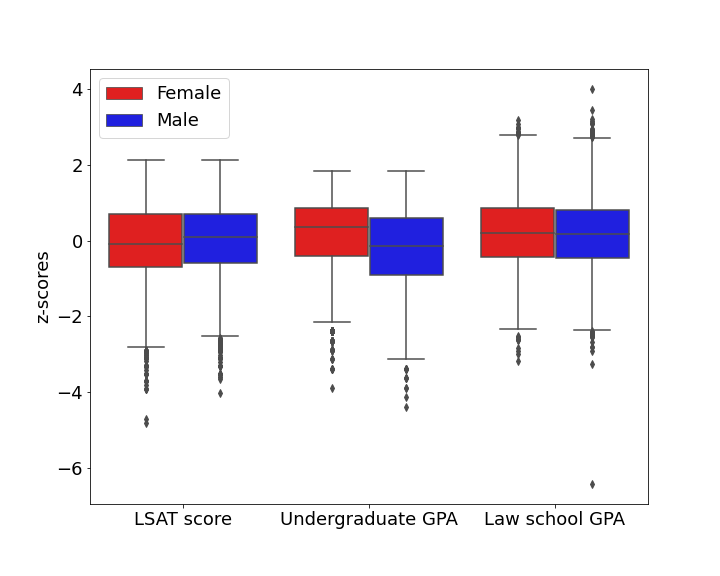}
}
\end{figure}

\begin{table}[!htbp]
\centering
\small
\caption{Trade-off between race parity and selected objective measures of meritocracy (reported in terms of $z$-scores of attributes: Law school GPA, Undergraduate GPA and LSAT score) in the LSAC dataset that is $\alpha$-biased with $\alpha = 0.3034$.  DP-SVM, DP-LR  achieve $\epsilon$-demographic parity for race with $\epsilon = 0.01$.
 }
\begin{tabular}{c|ccccc|ccccc}
\multicolumn{1}{c}{}
            &   \multicolumn{5}{c}{White subpopulation} & \multicolumn{5}{c}{Black subpopulation} \\
 Attribute  &  Data & SVM & DP-SVM & LR & DP-LR  & Data & SVM & DP-SVM & LR & DP-LR \\
    \hline
    Law school GPA (z-score)    & 0.24   & 0.24 & 0.28 & 0.25 & 0.28 & -0.74 & -0.75   & -0.90 & -0.74   & -0.89   \\ 
    UG GPA   & 3.27 & 3.27 & 3.28 & 3.27 & 3.28 & 2.94 & 2.94 & 2.91 & 2.94 & 2.92   \\
    LSAT score    & 37.78 & 37.79 & 37.90 & 37.78 & 37.92 & 30.90 & 30.87   & 30.21 & 30.92   & 30.27
    \end{tabular}
\label{tab:LSAC-metrics}
\end{table}

In Figure \ref{fig:LSAC-discrepancy}, we illustrate the discrepancy in distributions of LSAT scores,  law school GPA and university GPA of law school students who passed the bar exam grouped by race and gender labels. We observe significant differences between the distributions of GPA and LSAT scores across white and black subpopulations.  However, across gender labels we do not see any significant difference in the distribution of the scores among students who passed the bar exam except for undergraduate GPA which is slightly higher for female students. 

We employ Algorithm \ref{alg:debiasing} to train logistic regression and SVM classifiers that achieve $\epsilon$-demographic parity for both race and gender labels by placing meritocracy constraints on LSAT scores,  undergraduate GPA and law school GPA. In Table \ref{tab:LSAC-metrics}, we present a trade-off between LR,  SVM and their counterparts from Algorithm \ref{alg:debiasing} (DP-LR and DP-SVM) that achieve $\epsilon$-demographic parity (with $\epsilon = 0.01$) with respect to race.  We report the mean of some selected objective measures of meritocracy (Law school GPA, Undergraduate GPA and LSAT scores) in the LSAC dataset. 

We observe that mean of the attributes for students with positive outcome labels (using the output of the resulting models) in the white subpopulation does not change significantly even after employing Algorithm \ref{alg:debiasing} to achieve demographic parity.  However, the mean of the attributes for the positively labeled students in the black subpopulation decreases slightly signifying a lowering of threshold for passing the bar exam for black students that helps to achieve demographic parity.  For example, while in the white subpopulation we improve Law school GPA, UG GPA and LSAT scores slightly from $(0.24, 3.27, 37.78)$ in the dataset to $(0.28, 3.28, 37.92)$ for DP-LR,  for the black subpopulation we worsen from $(-0.74, 2.94, 30.90)$ to $(-0.89, 2.92, 30.27)$.  This is natural as we decrease the rate of positive observations in the white subpopulation and increase the rate in the black subpopulation.  In this sense,  and for this example,  the price of diversity is low. 

\begin{figure}[!htbp]
\centering
\caption{Comparison of trade-off between the disparity tolerance ($\epsilon$) vs.  change in the mean z-scores of attributes of interest with respect to protected variables (race and gender) using Algorithm \ref{alg:debiasing} (DP-LR) on LSAC dataset.  The y-axis denotes change in the z-scores of attributes, i.e.,  $\delta$-standard deviations in the z-score of an attribute.  For a given attribute,  a positive $\delta$ value signifies an increase in the mean value and similarly,  a negative $\delta$ value signifies a decrease in the mean value of that attribute in the modified dataset.  The attributes are LSAT scores (LSAT), undergraduate GPA (UGPA),  cumulative law school GPA (ZGPA) and first year law school GPA (ZFYGPA)}
\label{fig:epsilon-delta-tradeoff-lawschool}
\subfloat[DP-LR, sensitive attribute: race]{
\centering
\includegraphics[width=0.48\linewidth]{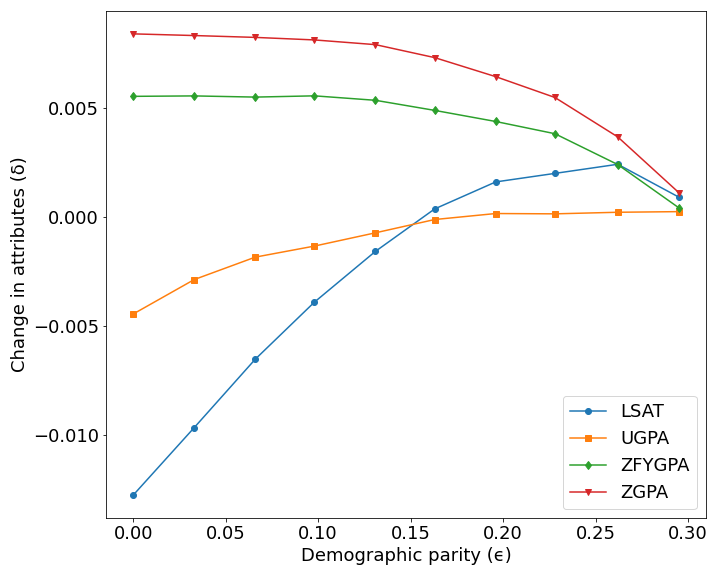}
}
\subfloat[DP-SVM, sensitive attribute: race]{
\centering
\includegraphics[width=0.48\linewidth]{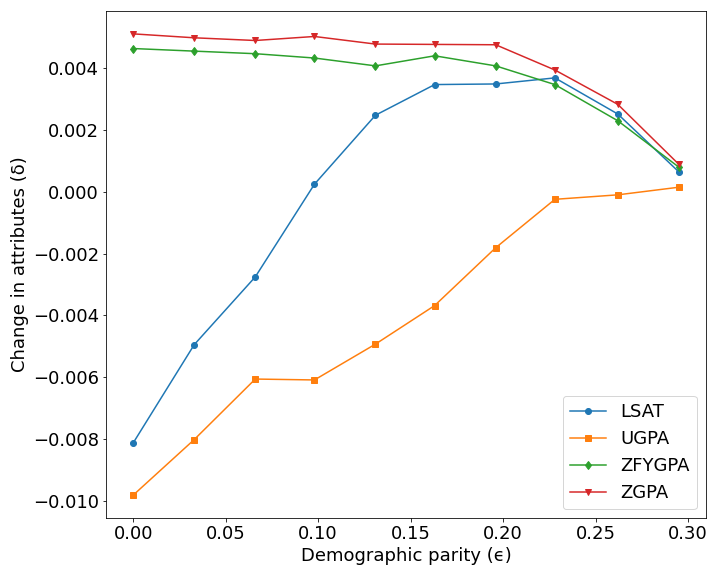}
}
\\
\subfloat[DP-LR, sensitive attribute: gender]{
\centering
\includegraphics[width=0.48\linewidth]{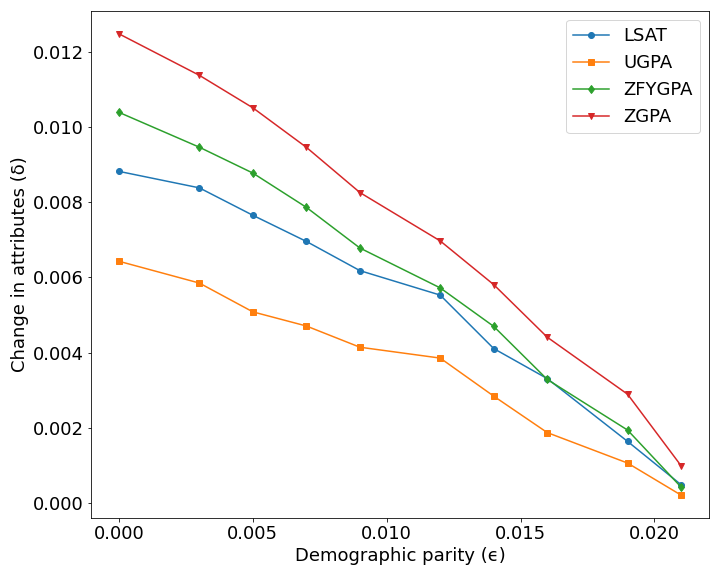}
}
\subfloat[DP-SVM, sensitive attribute: gender]{
\centering
\includegraphics[width=0.48\linewidth]{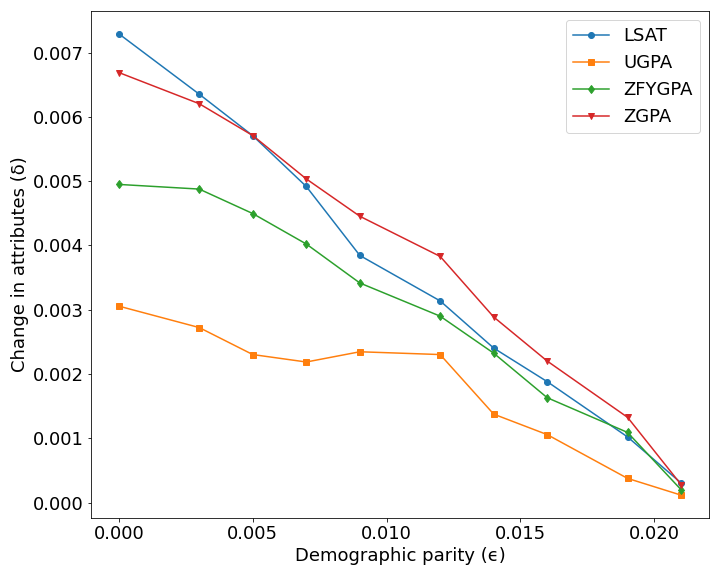}
}
\end{figure}

In Figure \ref{fig:epsilon-delta-tradeoff-lawschool},  we present trade-off curves between the demographic parity tolerance ($\epsilon$) and the change in the mean z-scores of attributes of interest (LSAT score, undergraduate GPA and law school GPA) using Algorithm \ref{alg:debiasing} for both race and gender labels.  In the trade-off curves using DP-LR for race labels (Figure \ref{fig:epsilon-delta-tradeoff-lawschool}(a)),  as we try to decrease demographic parity to near zero, the average merit measured in terms of LSAT scores and undergraduate GPA lowers by $0.013$ and $0.005$ standard deviations,  respectively compared to $0.01$ and $0.008$ using DP-SVM.  However,  average merit measured in terms of law school GPA (both cumulative GPA and first year GPA) increases by $(0.008, 0.005)$ standard deviations using DP-LR, and $(0.005, 0.004)$ using DP-SVM,  respectively.  In this case,  DP-LR model has a slight edge over DP-SVM in achieving a lower price for the attribute UGPA,  similar magnitude of price for LSAT scores,  and higher (cumulative and first year) lawschool GPA.

Interestingly,  for demographic parity across gender labels we observe that by trying to decrease the demographic parity tolerance,  we were able to increase the mean of all attributes (LSAT score, undergraduate GPA and law school GPA) of the students who pass the bar exam in the modified dataset on an average by $0.009$ and $0.005$ standard deviations using DP-LR and DP-SVM respectively (see Figures  \ref{fig:epsilon-delta-tradeoff-lawschool}(c,d)), i.e., the newly selected female students have higher scores compared to male students who they replace in the modified dataset.  Moreover,  the DP-LR model has superior performance in improving the mean of all attributes compared to DP-SVM model.  In this case the price of diversity with respect to gender is negative implying better outcomes among the students who pass the bar exam in the modified dataset using both DP-LR and DP-SVM.

\subsection{ProPublicas COMPAS}
In this second case study, we use data from Broward County, Florida originally compiled by ProPublica and published in their study \cite{angwin2016machine}. Following their analysis, we only consider black and white  defendants who were assigned COMPAS risk scores within a month of their arrest.  We use the COMPAS assigned risk scores (low/high) as an outcome to predict in this exercise and try to achieve demographic parity in those decisions using Algorithm \ref{alg:debiasing}.  Information about the defendents in the dataset includes age, race,  sex, number of prior convictions (priors count), and COMPAS violent crime risk scores. The resulting dataset is $\alpha$-biased with respect to race with$\alpha = 0.29$.

\begin{figure}[!htbp]
\centering
\caption{Distributions of age,  number of prior convictions and COMPAS decile scores for defendants who have been assigned a ``Low'' risk score grouped by race and gender in the COMPAS dataset, which is $\alpha$-biased with respect to race with $\alpha = 0.29$ and with respect to gender with $\alpha = 0.12$.}
\label{fig:COMPAS-discrepancy}
\subfloat[Distribution of attributes for race labels.]{
\centering
\includegraphics[width=0.49\linewidth]{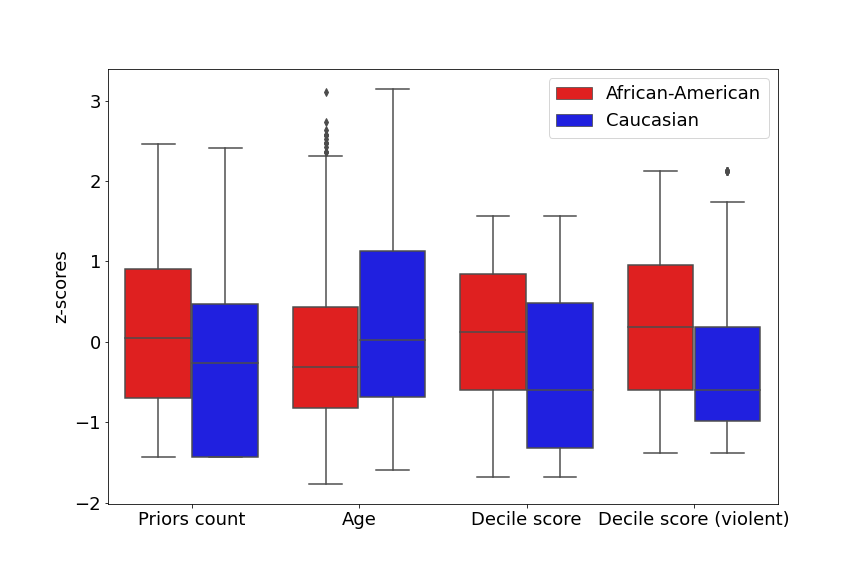}
}
\subfloat[Distribution of attributes for gender labels.]{
\centering
\includegraphics[width=0.49\linewidth]{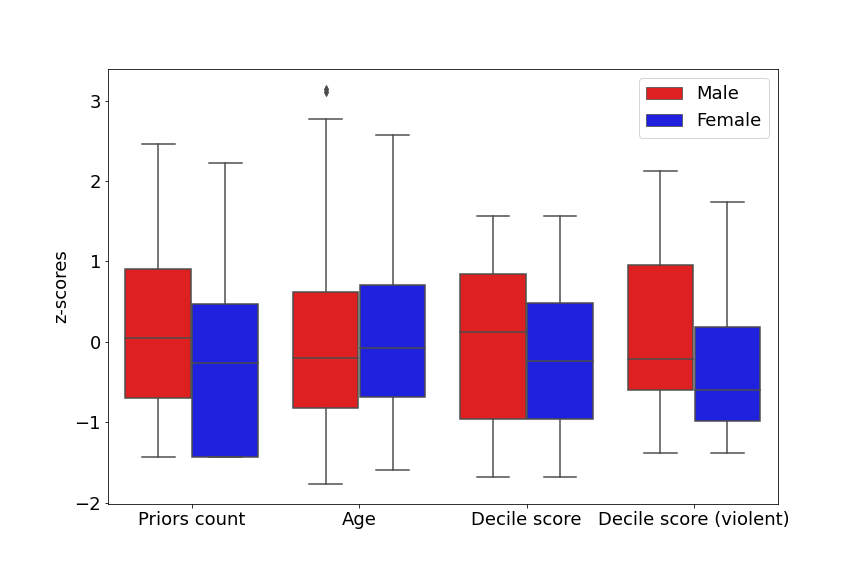}
}
\\
\subfloat[COMPAS decile scores for the white subpopulation.]{
\centering
\includegraphics[width=0.45\linewidth]{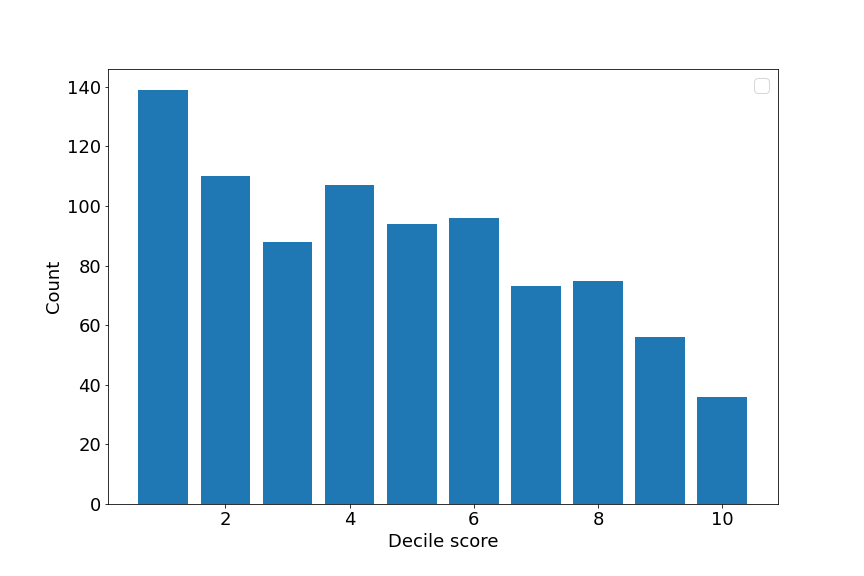}
}
\subfloat[COMPAS decile scores for the black subpopulation.]{
\centering
\includegraphics[width=0.45\linewidth]{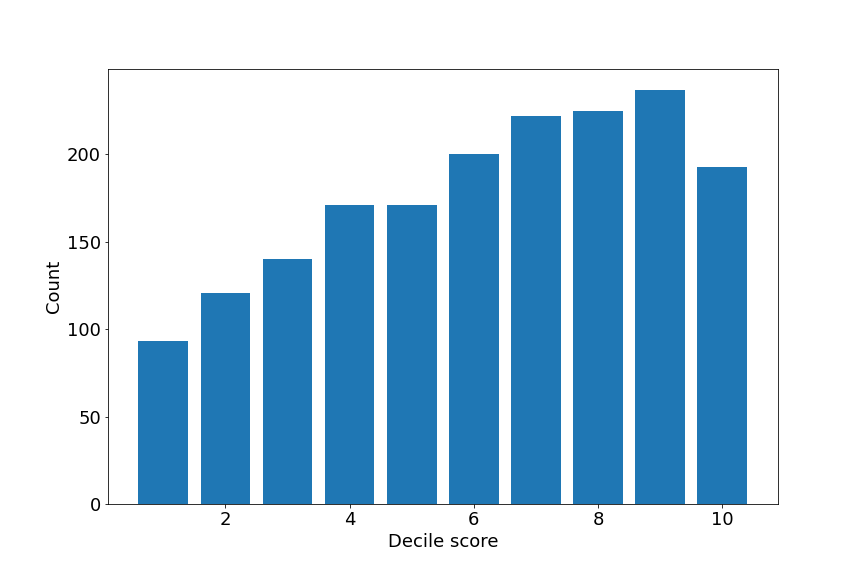}
}
\end{figure}

\begin{table}[!htbp]
\centering
\small
\caption{Trade-off between race parity and attributes of interest in the COMPAS dataset that is $\alpha$-biased with respect to race with  $\alpha = 0.29$.  DP-SVM, DP-LR achieve $\epsilon$-demographic parity with respect to race with  $\epsilon = 0.01$.
 }
\begin{tabular}{c|ccccc|ccccc}
\multicolumn{1}{c}{}
            &   \multicolumn{5}{c}{White  subpopulation} & \multicolumn{5}{c}{Black subpopulation} \\
 Metric   &   Data & SVM & DP-SVM & LR & DP-LR  &   Data & SVM & DP-SVM & LR & DP-LR          \\ 
    \hline
    Age   & 40.65 & 40.53  & 44.23   & 40.68 & 43.91  & 35.04 & 34.81   & 35.66 & 35.14   & 35.75  \\
    Decile score   & 2.08   & 2.14 & 1.81   & 2.02 & 1.87   & 2.45 & 2.38   & 3.22 & 2.41   & 3.28   \\
    Priors count    & 1.51   & 1.46 & 1.15   & 1.54 & 1.24   & 1.95 & 1.94   & 1.90 & 1.95   & 1.91   \\ 
    \end{tabular}
\label{tab:COMPAS-metrics}
\end{table}

In Figure \ref{fig:COMPAS-discrepancy}, we illustrate discrepancy in distributions of age, priors count and decile risk scores among defendants who received a ``Low'' risk score grouped by race and gender labels. We observe significant difference between the distributions across both race and gender labels: the black subpopulation were assigned a higher decile risk scores compared to the white  subpopulation,  and the male subpopulation were assigned a higher decile risk scores compared to the female subpopulation.  Also, the distributions of COMPAS decile scores are skewed towards 
 left (lower risk scores) for the white  subpopulation and skewed towards right (higher risk scores) for the black  subpopulation.

We employ Algorithm \ref{alg:debiasing} to train logistic regression and SVM classifiers that achieve $\epsilon$-demographic parity with respect to race with $\epsilon = 0.01$ by placing constraints on age,  decile scores and the number of prior convictions.  In Table \ref{tab:COMPAS-metrics}, we present a trade-off between vanilla LR, SVM and their counterparts DP-LR and DP-SVM that achieve $\epsilon$-demographic parity in terms of age,  decile scores and the number of prior convictions (prior counts). We observe that the mean age,  decile scores and the number of prior convictions of defendants who were assigned ``Low'' risk score in the original dataset changes from $(35.04, 2.45, 1.95)$ to $(35.75, 3.28,1.91)$ using predictions of DP-LR for the black  subpopulation.  Whereas for the white  subpopulation,  the mean age,  decile scores and the number of prior convictions of defendants who were assigned ``Low'' risk score in the original dataset changes from $(40.65,2.08,1.51)$ to $(43.91,1.87,1.24)$ for predictions of DP-LR.

\begin{figure}[!htbp]
\centering
\caption{Comparison of trade-off between the demographic parity tolerance ($\epsilon$) vs.  change in the mean z-scores of attributes of interest using Algorithm \ref{alg:debiasing} with respect to protected attribute: race on COMPAS dataset.  The y-axis denotes change in the z-scores of attributes, i.e.,  $\delta$-standard deviations in the z-score of an attribute.  For a given attribute,  a positive $\delta$ value signifies an increase in the mean value and similarly,  a negative $\delta$ value signifies a decrease in the mean value of that attribute in the modified dataset.  }
\label{fig:epsilon-delta-tradeoff-COMPAS}
\subfloat[DP-LR, sensitive attribute: race]{
\centering
\includegraphics[width=0.49\linewidth]{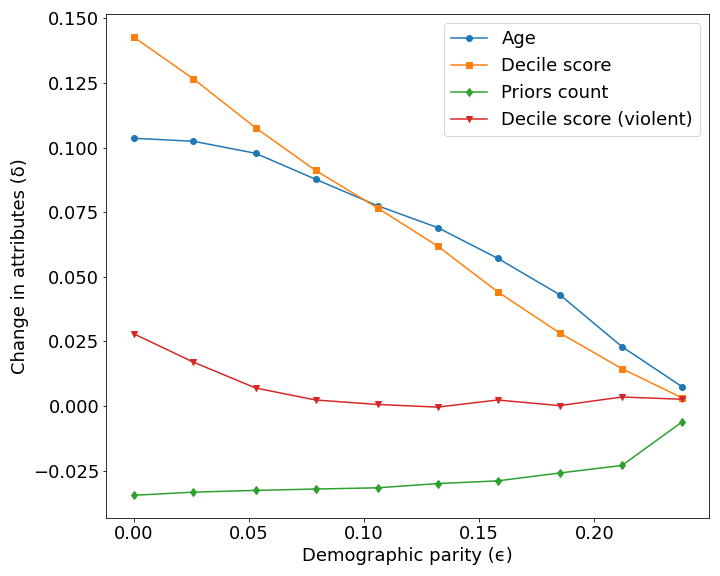}
}
\subfloat[DP-SVM, sensitive attribute: race]{
\centering
\includegraphics[width=0.49\linewidth]{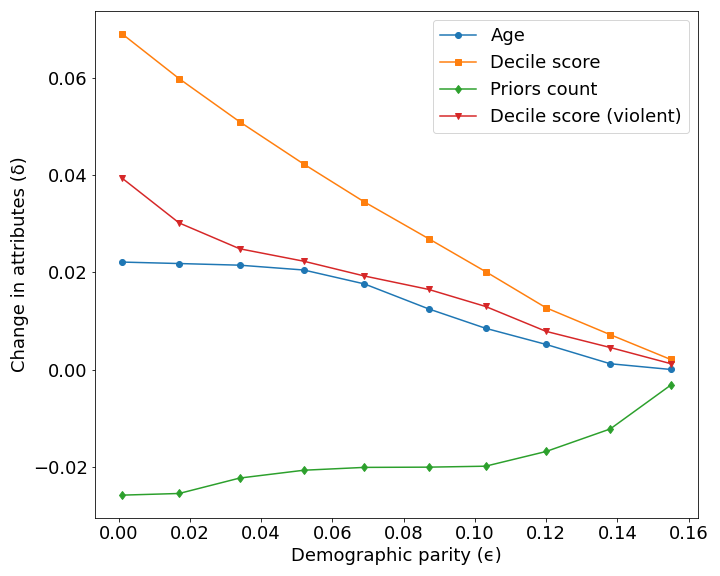}
}
\end{figure}

In Figure \ref{fig:epsilon-delta-tradeoff-COMPAS},  we present trade-off curves between the demographic parity tolerance ($\epsilon$) and the change in the mean z-scores of attributes of interest (age, decile score and priors count) using Algorithm \ref{alg:debiasing} for race labels.  As we try to decrease the demographic parity tolerance, the average number of prior counts for the defendants labeled as ``Low'' risk (in the modified dataset) decreases by similar magnitude using both DP-LR and DP-SVM models.  However,  average decile scores and age of defendants labeled as ``Low'' risk increases by $(0.14,  0.1)$ (using DP-LR) and $(0.06,  0.02)$ (using DP-SVM) standard deviations, respectively which signifies that defendants predicted to be ``Low'' risk have higher mean decile risk scores and are slightly older than individuals labeled as ``Low'' risk in the original dataset.  The average number of prior convictions decreases in both cases using DP-LR and DP-SVM by $0.03$ and $0.024$ standard deviations,  respectively.

Moreover,  the price of diversity for the attributes age and decile score using DP-SVM model is lower than DP-LR,  implying that DP-SVM model has superior performance in enhancing diversity with low price compared to DP-LR.  However,  the DP-LR model has an edge over DP-SVM in achieving a lower price for the attribute (violent) decile score ($0.026$ vs.  $0.04$), and improves (lowers) the average number of prior convictions by $0.03$ compared to $0.024$ using DP-SVM. This shows that the price of diversity for race in this example using both DP-LR and DP-SVM models is small (less than $0.15$ (DP-LR) and $0.067$ (DP-SVM) - standard deviations).

\subsection{Credit default dataset}
In this example,  we investigate whether there exists any bias in predicting whether an individual will default on their credit card payments for the subsequent month with respect to gender,  and try to alleviate such bias using Algorithm \ref{alg:debiasing}. The dataset consists of historical information about credit card users in Taiwan including  age,  gender, education, marital status, approved credit limit, history of payments,  bill and payment amounts.  For our analysis we convert approved credit limit,  average bill amount and average payment amount to log base 10.  Here,  the protected attribute is gender and the dataset is $\alpha$-biased with respect to gender with $\alpha = 0.033$.

\begin{table}[!htbp]
\centering
\small
\caption{Trade-off between race parity and attributes of interest in the credit default dataset that is $\alpha$-biased with respect to gender with $\alpha = 0.033$.  All reported values are log base 10 of the original numbers in the dataset.  DP-SVM, DP-LR achieve $\epsilon$-demographic parity for gender with $\epsilon = 0.001$.
 }
\begin{tabular}{c|ccccc|ccccc}
\multicolumn{1}{c}{}
            &   \multicolumn{5}{c}{Male subpopulation} & \multicolumn{5}{c}{Female subpopulation} \\
 Metric   &   Data & SVM & DP-SVM & LR & DP-LR  &   Data & SVM & DP-SVM & LR & DP-LR    \\ 
    \hline
    Credit approved  & 4.64 & 4.62  & 4.57   & 4.61 & 4.57  & 4.80 & 4.82  & 4.66 & 4.78   & 4.70 \\
    Bill amount  & 4.95   & 4.95 & 4.92   & 4.94 & 4.92   & 4.94 & 4.94   & 4.93 & 4.95   & 4.93  \\
    Payment amount  & 3.0   & 2.99 & 2.90   & 3.0 & 2.92   & 3.02 & 3.0   & 2.92 & 3.02 & 2.94 \\ 
    \end{tabular}
\label{tab:credit-default-metrics}
\end{table}

We employ Algorithm \ref{alg:debiasing} to train logistic regression and SVM classifiers that achieve $\epsilon$-demographic parity for both race and gender labels by placing constraints on approved credit limit, average bill and payment amounts. In Table \ref{tab:credit-default-metrics}, we present a trade-off between LR,  SVM and their counterparts from Algorithm \ref{alg:debiasing} (DP-LR and DP-SVM) that achieve $\epsilon$-demographic parity with $\epsilon = 0.001$.  We report the mean of various attributes of interest: approved credit line,  average bill amount and average payment amount in the credit default dataset. 

We observe that mean of the attributes for individuals with positive label (using the output of the resulting models) in both the male and female subpopulation do not change significantly even after employing Algorithm \ref{alg:debiasing} to achieve demographic parity.  For example,  while in the male subpopulation the mean of the attributes among individuals predicted to default decreases from $(4.64, 4.95, 3.0)$ in the dataset to $(4.57,4.92,2.92)$ for DP-LR for approved credit line, average bill amount and average payment amount respectively,  and similarly for the female subpopulation the mean of the attributes also decreases from $(4.8,4.94,3.02)$ to $(4.7,4.93,2.94)$. 

\begin{figure}[!htbp]
\centering
\caption{Comparison of trade-off between the demographic parity tolerance ($\epsilon$) vs.  change in the mean z-scores of attributes of interest using Algorithm \ref{alg:debiasing} with respect to protected attribute: gender on credit default dataset.  The y-axis denotes change in the z-scores of attributes, i.e.,  $\delta$-standard deviations in the z-score of an attribute.  For a given attribute,  a positive $\delta$ value signifies an increase in the mean value and similarly,  a negative $\delta$ value signifies a decrease in the mean value of that attribute in the modified dataset. }
\label{fig:epsilon-delta-tradeoff-credit-default}
\subfloat[DP-LR,  sensitive attribute: gender]{
\centering
\includegraphics[width=0.49\linewidth]{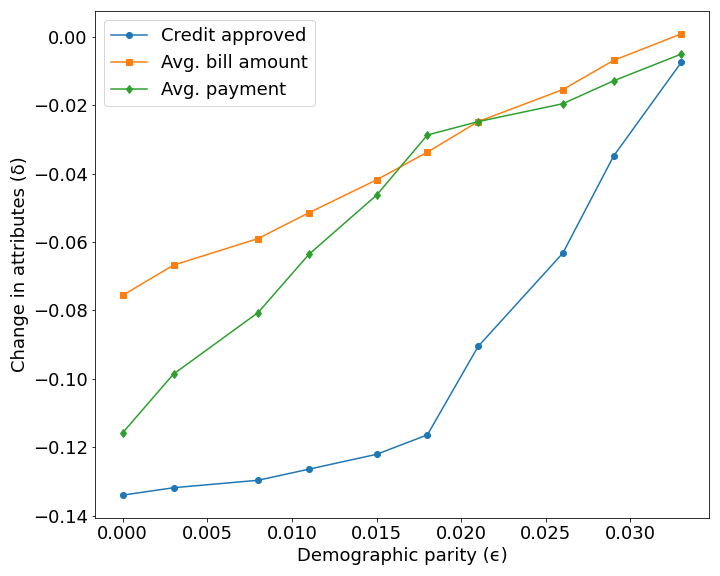}
}
\subfloat[DP-SVM,  sensitive attribute: gender]{
\centering
\includegraphics[width=0.49\linewidth]{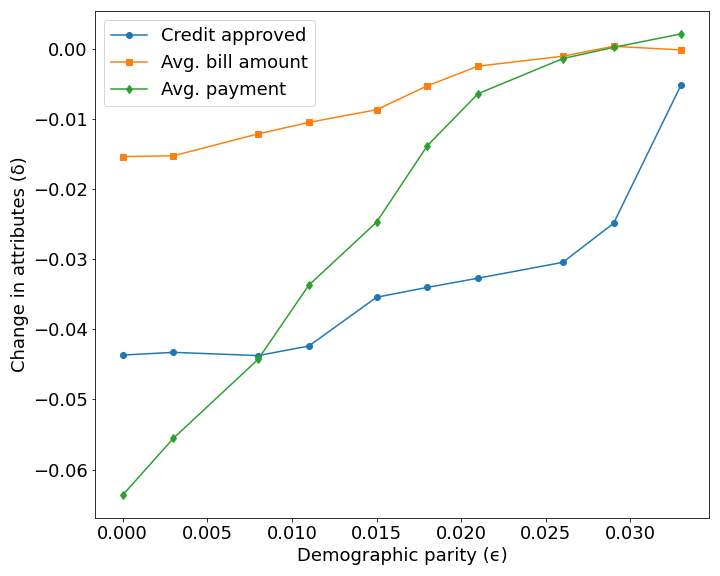}
}
\end{figure}

In Figure \ref{fig:epsilon-delta-tradeoff-credit-default},  we present trade-off curves between the demographic parity tolerance ($\epsilon$) and the change in the mean z-scores of attributes of interest (approved credit line, average bill amount and average payment amount) for gender labels using both DP-LR and DP-SVM models.  As we try to decrease the demographic parity tolerance,  the mean of all three attributes for individuals predicted to default on their payments decreases and this decrease is of the order of $[0.08,0.14]$ standard deviations using DP-LR and $[0.016,0.064]$ standard deviations using DP-SVM, i.e.,  in the modified dataset individuals labeled as high risk have lower lower approved credit line, average bill amount and average payment amount than individuals labeled as high risk in the original dataset.  Therefore,  DP-LR has superior performance in lowering the mean of all three attributes compared to DP-SVM.  Again, for this example the price of diversity with respect to gender is small: less than $(0.08,  0.12,  0.14)$ - standard deviations for bill amount, payment amount, approved credit limit, respectively.
\begin{table}[!htbp]
\centering
\small
\caption{Comparison of out-of-sample AUC with logistic regression (LR),  logistic regression with $\epsilon$-demographic parity (DP-LR),  vanilla SVM and SVM with $\epsilon$-demographic parity (DP-SVM).  DP-LR and DP-SVM were trained using Algorithm \ref{alg:debiasing} with respect to both race and gender labels with $\epsilon = 0.01$ for Law students dataset and only race labels for COMPAS dataset,  and with respect to gender labels with $\epsilon = 0.001$ for credit default dataset.
}
\begin{tabular}{l|c|c|c|c|}
 Dataset & LR & DP-LR & SVM & DP-SVM  \\ 
    \hline
     Law students (race)  & 0.867 & 0.856  & 0.850   & 0.851  \\
     Law students (gender) & 0.867 & 0.864  & 0.850   & 0.857 \\
     COMPAS  & 0.835   & 0.837 & 0.832  & 0.836   \\
     Credit default  & 0.740   & 0.748 & 0.763   & 0.767  \\ 
    \end{tabular}
\label{tab:AUC-summary}
\end{table}

In Table \ref{tab:AUC-summary},  we compare the performance of LR, SVM, DP-LR and DP-SVM models in terms of out-of-sample AUC on all three datasets.  For SVM models,  we observe a slight improvement in out-of-sample AUC using DP-SVM over vanilla SVM on all three datasets.  However,  for logistic regression, we do not observe a similar trend.  Overall, across the three case studies presented here,  we observe that the price of diversity is low and in some cases negative, leading to changes in selection processes that enhance diversity while also improving meritocracy.  In the next section, we propose an approach to implement changes to the current selection processes as discovered by Algorithm \ref{alg:debiasing} in an interpretable way.

\section{Implementation tool}\label{sec:implementation-tool}
In this section,  we use OCTs developed in  \citep{bertsimas2017optimal, MLbook}, which are highly interpretable and achieve state-of-the-art performance on classification problems to identify and differentiate individuals for whom the outcome label  is changed to either a \textsl{positive} or  \textsl{negative} label from individuals  with \textsl{no change} to the outcome labels.

In order to train an OCT for this purpose, we first construct a dataset based on the individuals for which we do/do not flip outcome labels using Algorithm \ref{alg:debiasing}. Each observation in the training dataset is labeled as one of the following: (a) \textsl{positive} (outcome label changed to a positive label),  (b) \textsl{negative} (outcome label changed to a negative label), or (c) \textsl{no change} (outcome label unchanged) based on the output of Algorithm \ref{alg:debiasing}. Using this dataset, we train a three-class OCT model with tree-depth chosen using cross-validation with depth: one through five for the ease of interpretability and select a model with the highest cross-validation accuracy among them. In Table \ref{tab:accuracy-epsilon-delta},  we present a summary of accuracy,  the difference between positive rates across gender/race labels ($\epsilon$) and maximum absolute change in the mean z-scores of meritocracy attributes ($\delta$) for OCT models on three test datasets.

\begin{table}[!htbp]
\centering
\small
\caption{Summary of OCT models with out-of-sample accuracy,  the difference between positive rates across gender/race labels ($\epsilon$) and maximum absolute change in the mean z-scores of meritocracy attributes ($\delta$) on test datasets.  DP-SVM, DP-LR were trained using Algorithm \ref{alg:debiasing} with respect to race labels with $\epsilon = 0.01$ for Law students and COMPAS datasets,  and with respect to gender with $\epsilon = 0.001$ for credit default dataset.
}
\begin{tabular}{l|c|c|c|c|}
 Dataset   &   OCT & Accuracy & $\epsilon$ & $\delta$  \\ 
    \hline
     \multirow{2}{*}{Law students}  & DP-LR & 0.94  & 0.011   & 0.014  \\
     & DP-SVM & 0.94  & 0.010   & 0.015  \\
     \multirow{2}{*}{COMPAS}  & DP-LR   & 0.80 & 0.016   & 0.122   \\
     & DP-SVM   & 0.82 & 0.014   & 0.117   \\
     \multirow{2}{*}{Credit default}  & DP-LR   & 0.84 & 0.002   & 0.136  \\ 
     & DP-SVM  & 0.89 & 0.001   & 0.133  \\ 
    \end{tabular}
\label{tab:accuracy-epsilon-delta}
\end{table}

\begin{figure}[!htbp]
\centering
\caption{Decision trees illustrating rules to achieve demographic parity for race in the law school students dataset. The attributes are age,  race (racetxt),  LSAT score (lsat) undergraduate GPA (ugpa),  cumulative law school GPA (zgpa) and first year law school GPA (zfygpa). }
\label{fig:tree-rules-LSAC-race}
\subfloat[An OCT model for predictions of DP-LR (out-of-sample accuracy: 0.94).]{
\centering
\includegraphics[width=\linewidth]{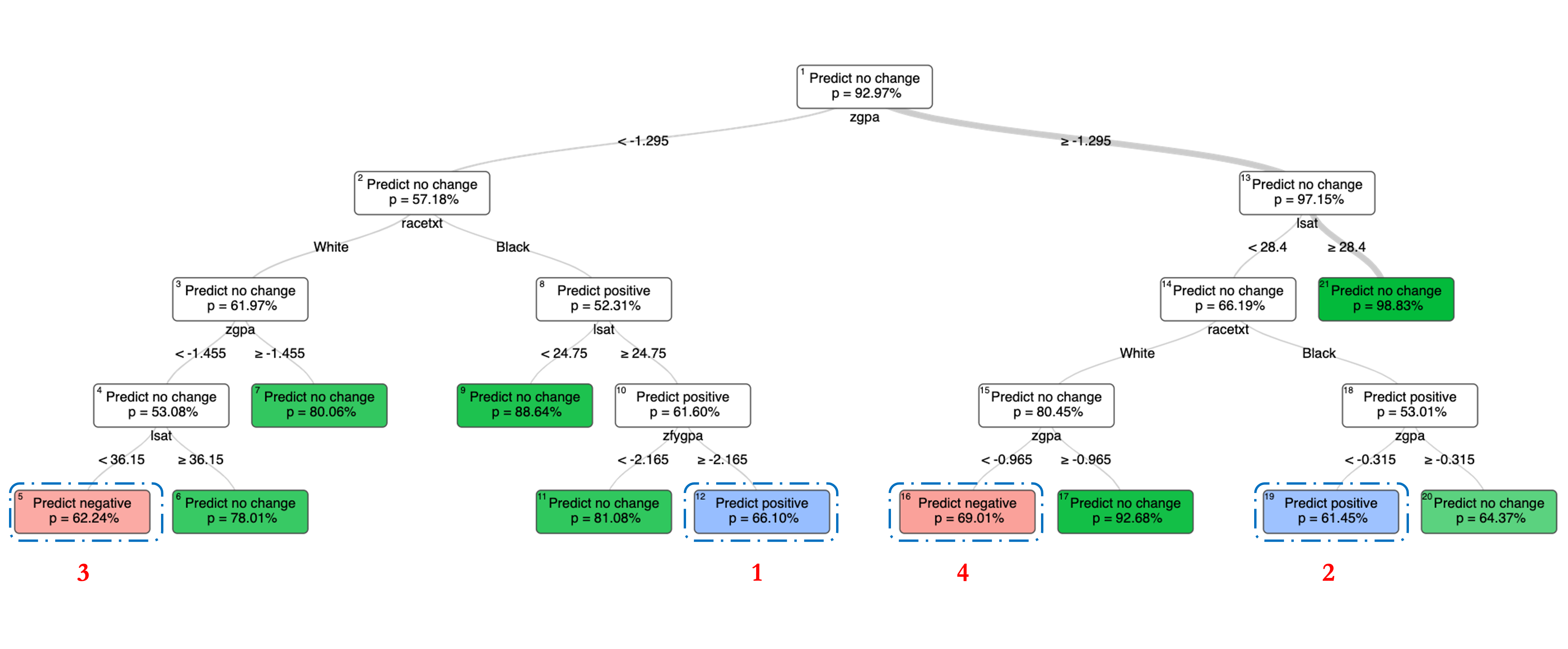}
}
\\
\subfloat[An OCT model for predictions of DP-SVM (out-of-sample accuracy: 0.94).]{
\centering
\includegraphics[width=\linewidth]{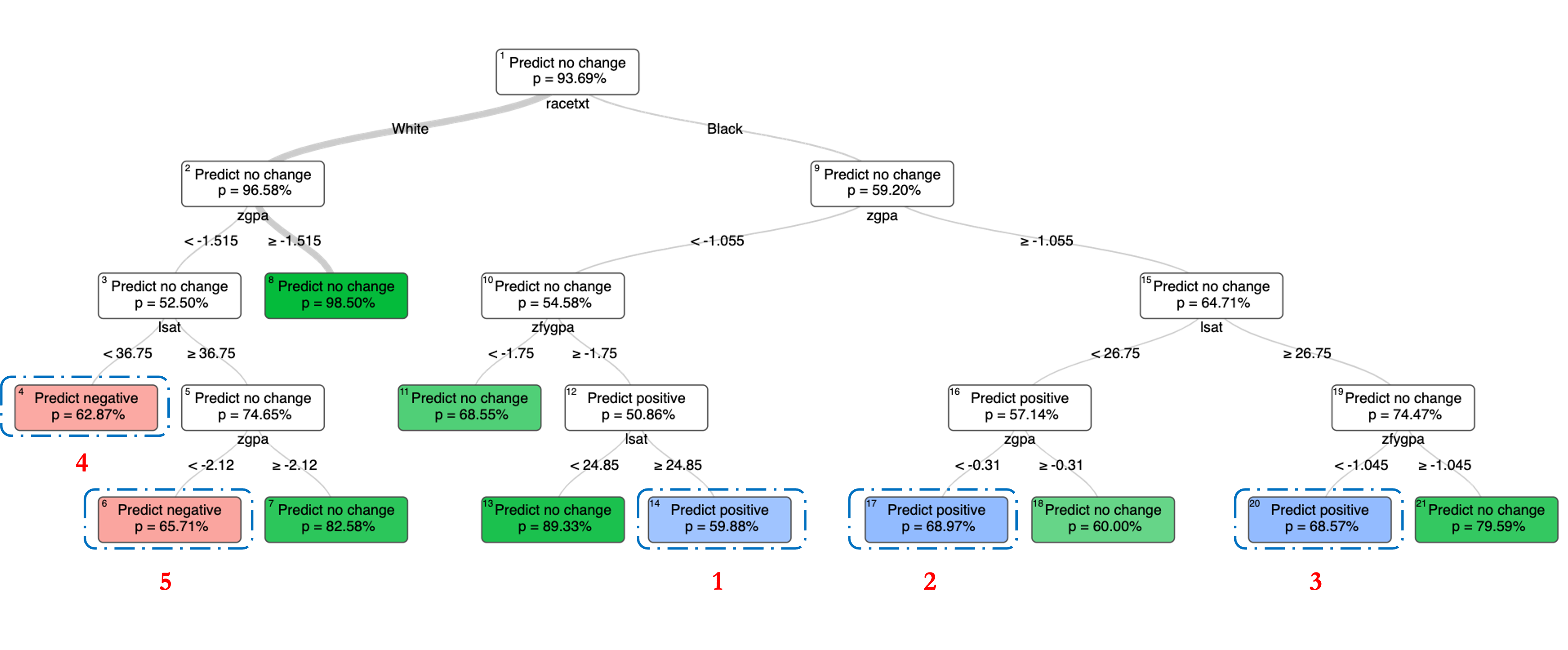}
}
\end{figure}

\begin{figure}[!htbp]
\centering
\caption{Decision trees illustrating rules to achieve demographic parity for race in the COMPAS dataset.  The attributes are age, race,  number of prior convictions (prior\_counts),  and number of days between screening and arrest (days\_b\_screening\_arrest). }
\label{fig:tree-rules-COMPAS-race}
\subfloat[An OCT model for predictions of DP-LR (out-of-sample accuracy: 0.80).]{
\centering
\includegraphics[width=6cm,height=19cm]{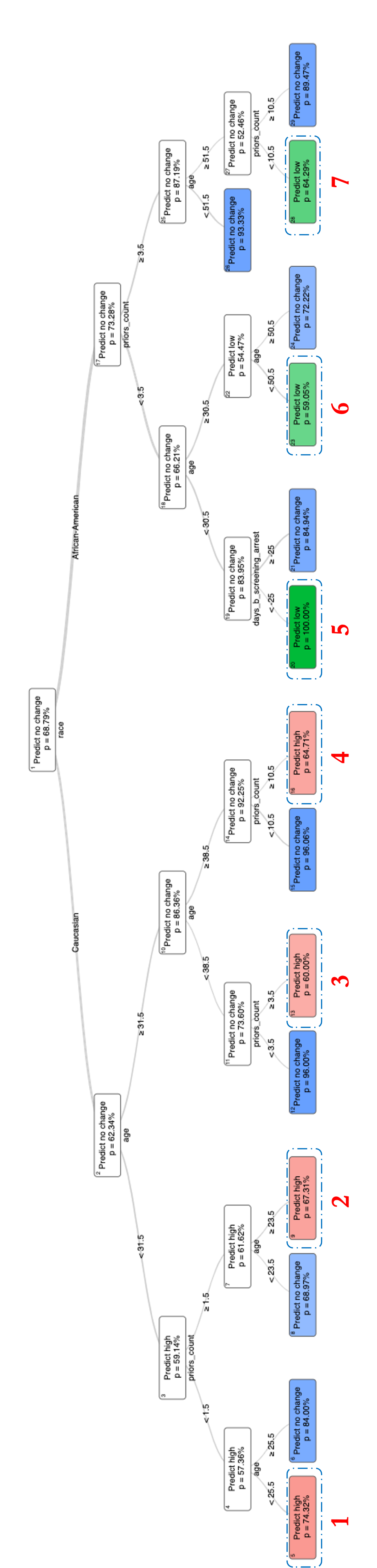}
}
\subfloat[An OCT model for predictions of DP-SVM (out-of-sample accuracy: 0.82).]{
\centering
\includegraphics[width=6cm,height=19cm]{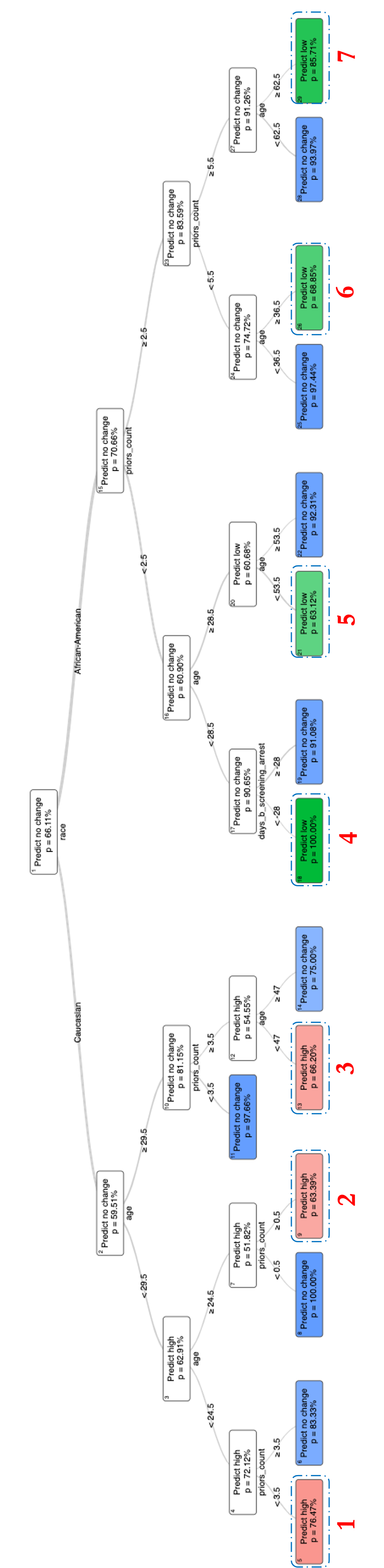}
}
\end{figure}


In Figures \ref{fig:tree-rules-LSAC-race}a and \ref{fig:tree-rules-LSAC-race}b, we present OCTs that approximate characteristics of law students for whom the outcome label was changed to either  a positive (passing the bar exam) or a  negative (failing the bar exam) label using DP-LR and DP-SVM models,  respectively.  Both OCT models identify and differentiate characteristics of the most meritorious black students that should be positively labeled, and the characteristics of white students for whom their outcomes should be flipped in order to achieve demographic parity and thereby improving diversity in the selected cohort.  


In Figure \ref{fig:tree-rules-LSAC-race}a,  among the black student subpopulation who originally failed the bar exam, DP-LR characterizes the most meritorious students based on the criteria approximated by the OCT model (Node 1: \{$\text{zgpa} < -1.295$ AND $\text{lsat} \geq 24.75$ AND $\text{zfygpa} \geq -2.165$\}  and   Node 2: \{$\text{zgpa} \geq -1.295$ AND $\text{lsat} < 28.4$ AND $\text{zgpa} < -0.315$\}),  and flips their outcome label to pass the bar exam.  In other words,  the algorithm selects black students with either strong cumulative GPA (zgpa) or stronger LSAT scores and first year GPA (zfygpa) to pass the bar exam.  Similarly,  among the white students subpopulation who originally passed the bar exam, DP-LR characterizes the least meritorious students based on the criteria as illustrated by the OCT model (Node 3: \{$\text{zgpa} < -1.455$ AND $\text{lsat} < 36.15$\}  and  Node 4: \{$\text{zgpa} \geq -1.295$ AND $\text{lsat} < 28.4$ AND $\text{zgpa} < -0.965$\}),  and flips their outcome label to fail the bar exam.  The algorithm selects white students with both lower cumulative GPA (zgpa) and LSAT scores with varying thresholds among students who originally passed the bar exam.

\begin{figure}[!htbp]
\centering
\caption{Decision trees illustrating rules to achieve demographic parity for race in the law school students dataset.  The attributes are age,  approved credit limit (LIMIT\_BAL), average bill amount (AVG\_BILL\_AMT) and average payment amount (AVG\_PAY\_AMT).}
\label{fig:tree-rules-credit-default-gender}
\subfloat[An OCT model for predictions of DP-LR (out-of-sample accuracy: 0.84).]{
\centering
\includegraphics[width=\linewidth]{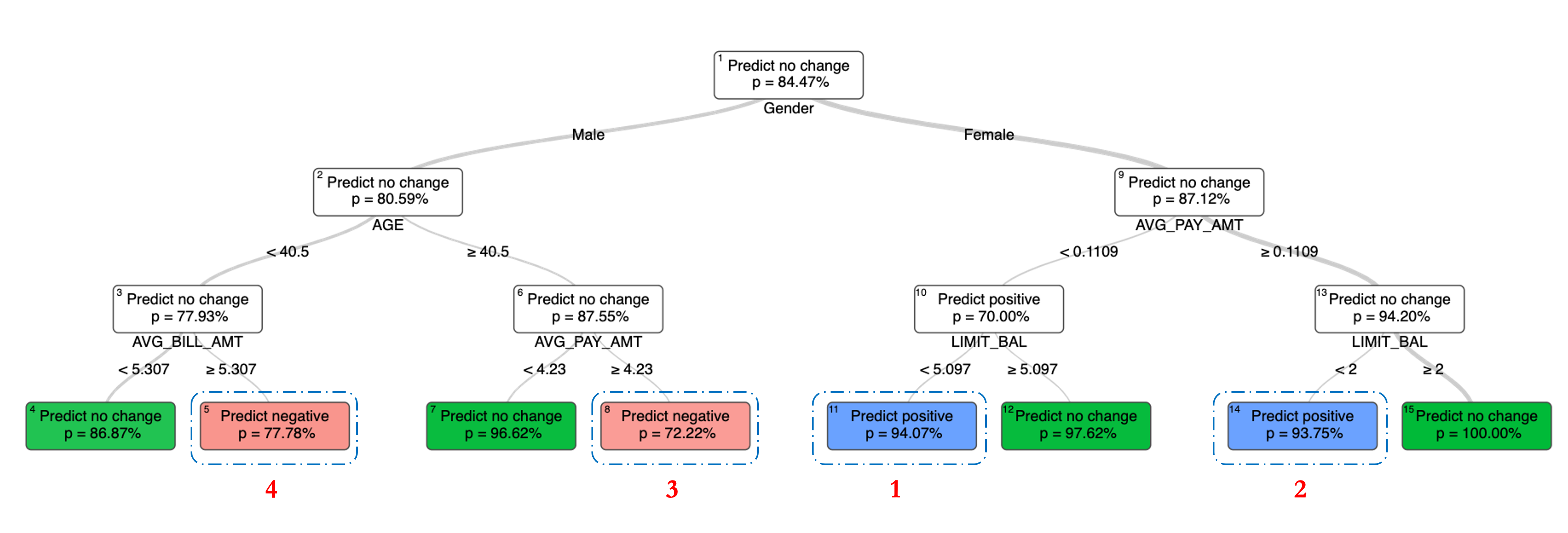}
}
\\
\subfloat[An OCT model for predictions of DP-SVM (out-of-sample accuracy: 0.89).]{
\centering
\includegraphics[width=\linewidth]{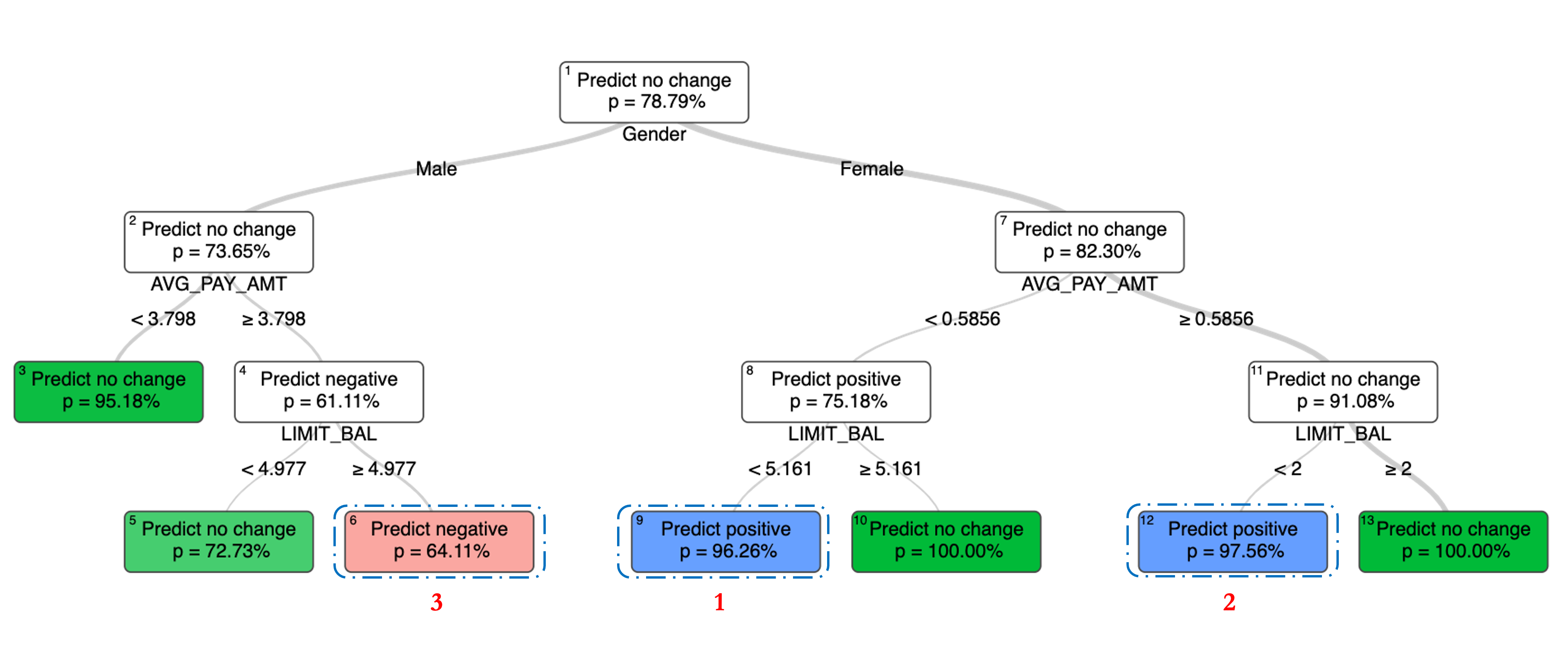}
}
\end{figure}

In Figures \ref{fig:tree-rules-COMPAS-race}a and \ref{fig:tree-rules-COMPAS-race}b,  we present OCTs  that differentiate characteristics of defendants for whom the outcome label was changed to either a positive (assigning a low risk score) or  a negative (assigning a high risk score) label using DP-LR and DP-SVM models, respectively on the COMPAS dataset. The OCT models approximate characteristics of black defendants who are most qualified to receive parole (using the outcomes predicted by DP-LR and DP-SVM) by being assigned a low risk score,  and the characteristics of white defendants who should be given a higher risk score in order to achieve demographic parity with respect to race and normalize distribution of risk score across the subgroups.  

In Figure \ref{fig:tree-rules-COMPAS-race}a,  among defendants in the African-American subpopulation,  DP-LR identifies criteria for low risk defendants approximated by the OCT model (Node 5: \{$\text{age} \leq 30$ AND $\text{prior counts} \leq 3$ AND $\text{days\_b\_screening\_arrest} < -25$\} and Node 6: \{$31 \leq \text{age} \leq 50$ AND $\text{prior counts} < 3$\} and Node 7: \{$\text{age} \geq 52$ AND $4 \leq \text{prior counts} \leq 10$\}) who were originally identified as high risk defendants in the COMPAS dataset.  In other words, the algorithm selects black defendants who are either young (with screening having taken place at least 25 days before arrest) or middle-aged with low number of prior convictions,  and relatively older defendants with moderately high number of prior convictions as low risk individuals.  Similarly,  among defendants in the white subpopulation, DP-LR identifies criteria for high risk individuals as illustrated in the OCT model (Node 1: \{$\text{age} \leq 25$ AND $\text{prior counts} \leq 1$\} and Node 2: \{$24 \leq \text{age} \leq 31$ AND $\text{prior counts} \geq 2$\} and Node 3: \{$32 \leq \text{age} \leq 38$ AND $\text{prior counts} \geq 4$\} and Node 4: \{$\text{age} \geq 39$ AND $\text{prior counts} \geq 11$\}),  and changed their outcome label as high risk defendants to achieve demographic parity.  The algorithm selects white defendants with relatively high number of prior convictions based on four different age groups ($0 - 25,  24 - 31, 32 - 38,  39+$) as high risk defendants.

Finally,  in Figures \ref{fig:tree-rules-credit-default-gender}a and \ref{fig:tree-rules-credit-default-gender}b,  we present OCTs that differentiate characteristics of individuals for whom the outcome label was changed to either a positive (default on credit card payment the following month) or a negative (not defaulting on payment) label using DP-LR and DP-SVM models, respectively on the credit default dataset. The OCT models approximate characteristics of women who received a negative label in the dataset and are more likely to default on payments (using the outcomes predicted by DP-LR and DP-SVM),  and the characteristics of men who received a positive label in the dataset and are least likely to default. In both of these cases, the OCT models illustrate characteristics of individuals for whom the labels are flipped in order to achieve demographic parity with respect to gender.  

In Figure \ref{fig:tree-rules-credit-default-gender}a,  among men who were expected to default on their credit card payments the subsequent month, the DP-LR model identifies criteria for individuals with lower risk to default as illustrated by the OCT model (Node 4: \{$\text{AGE} \leq 40$ AND $\text{AVG\_BILL\_AMT} \geq 5.307$\} and Node 3: \{$\text{AGE} \geq 40$ and $\text{AVG\_PAY\_AMT} \geq 4.23$\}).  In other words,  the algorithm selects either younger men with average bill amount greater than a certain amount or older men with average payment amount greater than a certain amount as low risk individuals.  Similarly,  among women who were not expected to default on their credit card payments the subsequent month, the DP-LR model identifies criteria (Node 1: \{$\text{AVG\_PAY\_AMT} < 0.1109$ AND $\text{LIMIT\_BAL} < 5.097$\}  and Node 2: \{$\text{AVG\_PAY\_AMT} < 4.23$ AND $\text{LIMIT\_BAL} < 2.0$\}),  and changed their outcome label as high risk individuals to achieve demographic parity.  The algorithm selects women with lower approved credit limit and/or lower average payment amount as high risk individuals to default on credit card payments for the subsequent month.

All six OCT models presented above approximate characteristics of individuals for whom the outcome labels are flipped so as to achieve demographic parity with respect to either race or gender labels. The OCT models should be used as an effective tool to implement changes in the current selection processes in a manner understandable by human decision makers to enhance diversity in the selected cohort (as discovered by the Algorithm \ref{alg:debiasing}) in a sytematic way.

\section{Conclusions}\label{sec:conclusions}
The key findings of this paper is that across the three real-world datasets the price of diversity is small,  that is we can modify the selection processes so as to enhance diversity without significantly affecting meritocracy.  We achieve this by using a novel optimization approach to train classification models with constraints on some selected objective measures of meritocracy to discover how to change the selection processes.  Furthermore, we propose a novel implementation tool employing OCTs to make changes to the current selection processes in a way that is understandable by human decision makers.  We believe that the methodology proposed in this paper contributes to alleviating bias and enhancing diversity in an interpretable and equitable way.

\bibliographystyle{abbrvnat}
\bibliography{refs}

\end{document}